\documentclass[aps,prb,preprint,showpacs,superscriptaddress]{revtex4}
\usepackage{times}
\usepackage{epsfig}
\usepackage{graphicx}
\usepackage{amsmath}
\begin{document}
\title{Multiple Scattering Theory for Two-dimensional Electron Gases in the Presence of Spin-Orbit Coupling}
\author{Jamie D. Walls}\affiliation{Department of Chemistry and Chemical Biology,
 Harvard University, Cambridge, MA 02138 USA}
 \author{Jian Huang}\affiliation{Department of Physics, Harvard
 University, Cambridge, MA 02138 USA}
 \author{Robert M. Westervelt}
 \affiliation{Department of Physics, Harvard University, Cambridge,
 MA 02138}\affiliation{Division of Engineering and Applied Sciences,
 Harvard University, Cambridge, MA 02138 USA}
\author{Eric J. Heller}
\email[Corresponding author: ]{heller@physics.harvard.edu}
\affiliation{Department of Chemistry and Chemical Biology, Harvard
University, Cambridge, MA 02138 USA} \affiliation{Department of
Physics, Harvard University, Cambridge, MA 02138 USA}
\date{\today}
\begin{abstract}
In order to model the phase-coherent scattering of electrons in
two-dimensional electron gases in the presence of Rashba spin-orbit
coupling, a general partial-wave expansion is developed for
scattering from a cylindrically symmetric potential.  The theory is
applied to possible electron flow imaging experiments using a
moveable scanning probe microscope tip. In such experiments, it is
demonstrated theoretically that the Rashba spin-orbit coupling can
give rise to spin interference effects, even for unpolarized
electrons at nonzero temperature and no magnetic field.
\end{abstract}
 \maketitle
\section{Introduction}
There has been recent interest\cite{Wolf01,Awschalombook,Zutic04} in
utilizing
 the spin degree of freedom in semiconductor devices, where
the charge carrier's spin provides an additional degree of control
and flexibility towards developing devices that are faster and more
efficient devices than conventional electronic devices.  One
component of potential ``spintronic" devices, the spin transistor
proposed by Datta and Das\cite{Datta90}, modulates the current
passing through a semiconductor due to the presence of the
spin-orbit interaction, which couples the electron's spin with its
kinematical motion. Interest in the spin transistor has generated
numerous theoretical and experimental investigations into the spin
dynamics under the spin-orbit interaction in two-dimensional
electron gases (2DEG).

 In layered semiconductors
devices, the two predominant sources of spin-orbit coupling arise
from either structure inversion asymmetry (SIA or Rashba
interaction\cite{Bychkov84}) or  bulk inversion asymmetry (BIA or
Dresselhaus interaction\cite{Dresselhaus55}).  The BIA spin-orbit
interaction arises from the breaking of inversion symmetry by the
inherent asymmetry of the atomic arrangement in the structure and is
not very amenable to external manipulation. The Rashba spin-orbit
coupling, on the other hand, arises from band bending at the
interfaces between semiconductor layers and/or any external electric
fields applied to the the device. Unlike the Dresselhaus coupling,
the strength of the Rashba coupling can be partially controlled by
application of an external electric field\cite{Nitta97} and in
principle can be made the dominant form of spin-orbit interaction in
the 2DEG. Such tunability of the Rashba interaction is ideally
suited for applications in spintronic devices, and as such, only the
Rashba spin-orbit coupling will be considered in this study.

Numerous studies have been conducted on the diffusive transport of
spins in the presence of spin-orbit
coupling\cite{Mishchenko04,Burkov04,Schliemann03} in order to
investigate a variety of phenomena, such as the spin Hall
effect\cite{Hirsch99,Zhang00}. Most of the studies were conducted up
to the first-Born approximation for the scattering from nonmagnetic
impurities, and the results were disorder averaged.  However, there
are many cases where such statistical theories are not warranted.
 For example, coherent scattering from a fixed set of impurities, which give rise
 to quantum
 interference effects induced by multiple-scattering events
from the localized impurities, can't be described by such
statistical theories. One method of tackling such problems is
multiple scattering theory, which has been routinely used in optical
and acoustic scattering and has been proposed as a method for
understanding the fringing patterns in recent imaging experiments on
electron flow in 2DEG\cite{Topinka01,shawpap}. In scattering theory,
the effect of a scatterer $k$ can be localized to a point in space
at the center of the scatterer, $\vec{r}_{k}$, such that an
operator, $\widehat{T}_{k}$, can be constructed which generates the
scattered wave, $\Psi_{S}(\vec{R})$, from the incident wave,
$\Phi_{\text{in}}(\vec{R})$, evaluated at the site of the scatterer,
$\vec{R}=\vec{r}_{k}$:
\begin{eqnarray} \Psi(\vec{R})&=&\Phi_{\text{in}}(\vec{R})+\Psi_{S}(\vec{R})\nonumber\\
&=&\Phi_{\text{in}}(
\vec{R})+\left(\widehat{T}_{k}(\vec{R},\vec{r})\Phi_{\text{in}}(\vec{r})\right)_{\vec{r}=\vec{r}_{k}}
\label{eq:tintro}\end{eqnarray} The subscript, $\vec{r}=\vec{r}_{k}$
means to operate $\widehat{T}_{k}(\vec{R},\vec{r})$ upon
$\Phi_{\text{in}}(\vec{r})$ and evaluate the result at
$\vec{r}=\vec{r}_{k}$. In the presence of N point scatterers, the
total wave function is then given by
\begin{eqnarray}
\Psi(\vec{R})&=&\Phi_{\text{in}}(\vec{R})+\sum_{k=1}^{N}\left(\widehat{T}_{k}(\vec{R},\vec{r})\Psi(\vec{r})\right)_{\vec{r}=\vec{r}_{k}}
\end{eqnarray}
Thus the complete wave function can be found if
$\left(\widehat{T}_{k}(\vec{R},\vec{r})\Psi(\vec{r})\right)_{\vec{r}=\vec{r}_{k}}$
is known at each scatter $k$.

In the following article, a multiple-scattering theory in the
presence of Rashba spin-orbit coupling in a 2DEG is developed. The
general formalism is presented, along with the explicit calculation
of the scattering operator, $\widehat{T}_{k}$, for a cylindrically
symmetric well/barrier, which will be used as a model for impurities
in a 2DEG.  As an application, the methodology is applied to
possible flux measurements for phase-coherent transport in a 2DEG
with Rashba spin-orbit interaction in the presence of a scanning
probe microscope (SPM) tip in zero magnetic field.  Additional
interference effects arise in the flux measurements due to spin
interference effects caused by the Rashba coupling.

\section{Scattering from a cylindrically symmetric potential:  Partial-wave expansion}
 The Hamiltonian for a 2DEG in the
presence of the Rashba spin-orbit interaction and impurities is
given by
\begin{eqnarray}
\widehat{H}&=&\frac{\widehat{p}_{X}^{2}}{2m^{*}}+\frac{\widehat{p}_{Y}^{2}}{2m^{*}}-\frac{\alpha}{\hbar}\left(\widehat{p}_{Y}\widehat{\sigma}_{X}-\widehat{p}_{X}\widehat{\sigma}_{Y}\right)+V(x,y)\nonumber\\
&=&\widehat{H}_{0}+V(x,y) \label{eq:Ho}
\end{eqnarray}
where $\widehat{\sigma}_{j}$ are the Pauli spin matrices, and
$\alpha$ is the Rashba spin-orbit coupling constant.  The
eigenstates and corresponding eigenvalues for the free-particle
Hamiltonian with Rashba spin-orbit coupling, $\widehat{H}_{0}$,  are
given by
\begin{eqnarray}
|k(\theta),\pm(\theta)\rangle&=&|k_{Y}=k\cos(\theta),k_{X}=k\sin(\theta)\rangle|\pm(\theta)\rangle
\\
E_{\pm}&=&\frac{\hbar^{2}k^{2}}{2m^{*}}\mp\alpha k
\label{eq:disperso}\end{eqnarray} where \begin{eqnarray}
\tan(\theta)&=&\frac{k_{X}}{k_{Y}}\nonumber\\
|\pm(\theta)\rangle&=&\frac{1}{\sqrt{2}}\binom{1}{\pm
\exp(-\text{i}\theta)}
\end{eqnarray}
and $k=\sqrt{k^{2}_{X}+k^{2}_{Y}}$.

The dispersion relation in Eq.~(\ref{eq:disperso}) represents two
parabolic bands centered upon $k=\pm m^{*}\alpha/\hbar^{2}$. For
states propagating with their momentum vectors making an angle
$\theta$ with respect to the $\widehat{Y}$-axis and for an energy
$E\geq 0$, there exists a two-fold degeneracy with the degenerate
states given by
\begin{eqnarray}
\label{eq:k1}
 |k_{1}(\theta),+(\theta)\rangle&=&|
\vec{k}_{1}(\theta)\rangle\sqrt{\frac{1}{2}}\binom{1}{\exp(-\text{i}\theta)}\\
|k_{2}(\theta),-(\theta)\rangle&=&|\vec{k}_{2}(\theta)\rangle\sqrt{\frac{1}{2}}\binom{1}{-
\exp(-\text{i}\theta)}\label{eq:k2}
\end{eqnarray}
where
$\vec{k}_{1(2)}=k_{1(2)}(\cos(\theta)\widehat{Y}+\sin(\theta)\widehat{X})$
with
\begin{eqnarray}
k_{1}&=&\frac{m^{*}\alpha}{\hbar^{2}}+\sqrt{\left(\frac{m^{*}\alpha}{\hbar^{2}}\right)^{2}+\frac{2m^{*}E}{\hbar^{2}}}\nonumber\\
k_{2}&=&-\frac{m^{*}\alpha}{\hbar^{2}}+\sqrt{\left(\frac{m^{*}\alpha}{\hbar^{2}}\right)^{2}+\frac{2m^{*}E}{\hbar^{2}}}
\label{eq:defk}
\end{eqnarray}
The states $|k_{1}(\theta),+(\theta)\rangle$ and
$|k_{2}(\theta),-(\theta)\rangle$ represent plane-wave states whose
spin states are quantized in the plane, perpendicular to the
momentum direction.

 In polar coordinates, which are useful when
considering scattering from a localized, cylindrically symmetric
potential, $\widehat{H}_{0}$ can be written as
\begin{eqnarray}
\widehat{H}_{0}&=&-\frac{\hbar^{2}}{2m^{*}}\left(\frac{\partial^{2}}{\partial r^{2}}+\frac{1}{r}\frac{\partial}{\partial r}+\frac{1}{r^{2}}\frac{\partial^{2}}{\partial\theta^{2}}\right)+\text{i}\alpha\left(\begin{array}{cc}0&\exp(\text{i}\theta)\left(\frac{\partial}{\partial r}+\frac{\text{i}}{r}\frac{\partial}{\partial\theta}\right)\\
\exp(-\text{i}\theta)\left(\frac{\partial}{\partial
r}-\frac{\text{i}}{r}\frac{\partial}{\partial\theta}\right)&0\end{array}
\right)\nonumber\\
\label{eq:hpolo}
\end{eqnarray}
The eigenstates of $\widehat{H}_{0}$ which represent states
propagating outward from or towards a particular origin,
$\vec{r}_{i}$, can be written as \begin{eqnarray}
\langle \vec{R}|\chi^{\pm}_{l,\uparrow,E}\rangle=\chi^{\pm}_{l,\uparrow}(\vec{R},E)&=&\exp(\text{i}l\theta)\frac{\sqrt{k_{1}}}{2\sqrt{2}}\binom{H^{\pm}_{l}(k_{1}|\vec{R}-\vec{r}_{i}|)}{-\text{i}H^{\pm}_{l-1}(k_{1}|\vec{R}-\vec{r}_{i}|)\exp(-\text{i}\theta)}\nonumber\\
\langle \vec{R}|\chi^{\pm}_{l,\downarrow,E}\rangle=\chi^{\pm}_{l,\downarrow}(\vec{R},E)&=&\exp(\text{i}l\theta)\frac{\sqrt{k_{2}}}{2\sqrt{2}}\binom{H^{\pm}_{l}(k_{2}|\vec{R}-\vec{r}_{i}|)}{\text{i}H^{\pm}_{l-1}(k_{2}|\vec{R}-\vec{r}_{i}|)\exp(-\text{i}\theta)}\nonumber\\
\label{eq:polstates}
\end{eqnarray}
where $H^{\pm}_{l}(z)$ are Hankel functions given by
$H^{\pm}_{l}(z)=J_{l}(z)\pm \text{i}Y_{l}(z)$, and $k_{1}$ and
$k_{2}$ are given in Eq.~(\ref{eq:defk}).  A similar solution to
Eq.~(\ref{eq:hpolo}) for a cylindrical well has been given
before\cite{Bulgakov01,Cserti04}.  The states
$\chi^{\pm}_{l,\uparrow(\downarrow)}(\vec{R},E)$ satisfy a flux
orthogonality condition through a circular surface surrounding the
origin, $\vec{r}_{i}$, which is given by
\begin{eqnarray}\frac{1}{2}\int^{2\pi}_{0}\langle
\chi^{\pm}_{l,a,E}|\left(|\vec{R}(\theta)\rangle\langle\vec{R}(\theta)|\vec{\widehat{J}}+\vec{\widehat{J}}|\vec{R}(\theta)\rangle\langle\vec{R}(\theta)|\right)|\chi^{\pm}_{m,b,E}\rangle\cdot\vec{R}(\theta)\text{d}\theta&=&\frac{\hbar
\overline{k}}{m^{*}}\delta_{l,m}\delta_{a,b}\end{eqnarray} where the
current operator, $\vec{\widehat{J}}$, is given by
\begin{eqnarray}
\vec{\widehat{J}}&=&\widehat{J}_{X}\widehat{X}+\widehat{J}_{Y}\widehat{Y}\nonumber\\
&=&\left(\frac{\widehat{p}_{X}}{m^{*}}+\frac{\alpha}{\hbar}\widehat{\sigma}_{Y}\right)\widehat{X}+\left(\frac{\widehat{p}_{Y}}{m^{*}}-\frac{\alpha}{\hbar}\widehat{\sigma}_{X}\right)\widehat{Y}
\end{eqnarray}

\begin{figure}[tbp]
\centering \includegraphics[width=10.7cm]{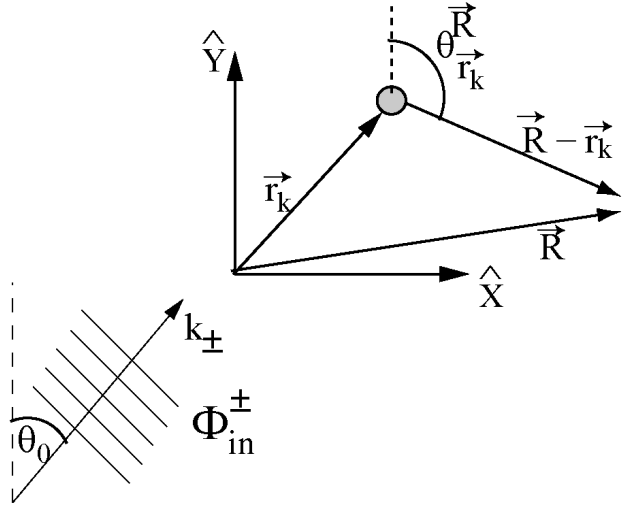}
\caption{Scattering of an incident plane wave
$\Phi^{\pm}_{\text{in}}(\vec{R})$ [Eq.~(\ref{eq:inwave})] from a
potential located at $\vec{r}_{k}$, $V_{k}(\vec{R})$
[Eq.~(\ref{eq:pot})]. The various angles and vectors used in
Eq.~(\ref{eq:inwave}) and Eq.~(\ref{eq:ang}) are illustrated.}
\label{fig:figure1}
\end{figure}

The states in Eq.~(\ref{eq:polstates}) can be used to generate the
scattering operator, $\widehat{T}_{k}$ [Eq.~(\ref{eq:tintro})], for
scattering in the presence of Rashba spin-orbit coupling.  The
following treatment follows closely a previous treatment for
constructing $\widehat{T}_{k}$ in the absence of spin-orbit
coupling.\cite{Herschthesis} We will begin by solving the
Schrodinger equation for an eigenstate of $\widehat{H}_{0}$
[Eq.~(\ref{eq:Ho})] incident upon a cylindrically symmetric
potential centered at $\vec{r}_{k}$, $V_{k}(\vec{r})$, as shown in
Figure \ref{fig:figure1}.   $V_{k}(\vec{r})$ is given by
\begin{eqnarray}
V_{k}(\vec{r})&=&\begin{array}{ccc}V_{0}\,\,&\text{for}&|\vec{r}-\vec{r}_{k}|\leq a\\
0\,\,&\text{for}&|\vec{r}-\vec{r}_{k}|>a\end{array}\label{eq:pot}\end{eqnarray}
where $a$ is the radius of the scattering potential.  Consider the
case of an incident wave propagating with momentum
$\vec{k}_{\pm}=k_{\pm}(\cos(\theta_{0})\widehat{Y}+\sin(\theta_{0})\widehat{X})$,
where $\pm$ denotes a particular eigenstate of $H_{0}$
[$|\vec{k}_{+}(\theta_{0}),+(\theta_{0})\rangle=|\vec{k}_{1}(
\theta_{0}),+(\theta_{0})\rangle$ and
$|\vec{k}_{-}(\theta_{0}),-(\theta_{0})\rangle=|\vec{k}_{2}(
\theta_{0}),-(\theta_{0})\rangle$ given in Eqs.
(\ref{eq:k1})-(\ref{eq:k2})]. The incident wave function,
$\Phi_{\text{in}}^{\pm}(\vec{R})$, written in a coordinate system
centered about the scatterer at $\vec{r}_{k}$ is given by
\begin{eqnarray}
\Phi^{\pm}_{\text{in}}(\vec{R})&=&\frac{1}{\sqrt{2}}\exp\left(\text{i}\vec{k}_{\pm}\cdot\vec{R}\right)\binom{1}{\pm
\exp(-\text{i}\theta_{0})}\nonumber\\
&=&\frac{1}{\sqrt{2}}\exp(\text{i}\vec{k}_{\pm}\cdot\vec{r}_{k})\exp(\text{i}k_{\pm}r_{\vec{R},\vec{r}_{k}}\cos(\theta_{0}-\theta^{\vec{R}}_{\vec{r}_{k}}))\binom{1}{\pm\exp(-\text{i}\theta_{0})}\nonumber\\
&=&\frac{1}{\sqrt{2}}\exp(\text{i}\vec{k}_{\pm}\cdot\vec{r}_{k})\binom{1}{\pm
\exp(-\text{i}\theta_{0})}\left(\sum_{l=-\infty}^{\infty}J_{l}(k_{\pm}r_{\vec{R},\vec{r}_{k}})\text{i}^{l}\exp\left(\text{i}l\left[\theta^{\vec{R}}_{\vec{r}_{k}}-\theta_{0}\right]\right)\right)
\label{eq:inwave}
\end{eqnarray}
where $r_{\vec{R},\vec{r}_{k}}=|\vec{R}-\vec{r}_{k}|$ is the
distance measured from the center of the scatterer and
$\theta^{\vec{R}}_{\vec{r}_{k}}$ is the angle with respect to the
$\widehat{Y}$-axis of the vector $\vec{r}$, i.e.,
\begin{eqnarray}
\exp\left(\text{i}\theta^{\vec{R}}_{\vec{r}_{k}}\right)&=&\frac{(\vec{R}-\vec{r}_{k})\cdot\widehat{Y}+\text{i}(\vec{R}-\vec{r}_{k})\cdot\widehat{X}}{r_{\vec{R},\vec{r}_{k}}}
\label{eq:ang}
\end{eqnarray}

The wave function outside of the scatterer,
$|\vec{R}-\vec{r}_{k}|>a$, can therefore be written as
\begin{eqnarray}
\Psi^{\pm}_{\text{I}}(\vec{R})&=&\Phi^{\pm}_{\text{in}}(\vec{R})+\Psi_{S}^{\pm}(\vec{R})
\end{eqnarray}
where the scattered wave function, $\Psi_{S}^{\pm}(\vec{R})$, can be
written as
\begin{eqnarray}
\Psi^{\pm}_{S}(\vec{R})&=&\sum_{l=-\infty}^{\infty}f^{\pm
1}_{l}\chi^{+}_{l,\uparrow}(\vec{R},E)+f^{\pm
2}_{l}\chi^{+}_{l,\downarrow}(\vec{R},E)
\end{eqnarray}
The wave function inside the cylindrical potential,
$\Psi^{\pm}_{\text{II}}(\vec{R})$, can be similarly written for
$|\vec{R}-\vec{r}_{k}|\leq a$ as
\begin{eqnarray}
\Psi^{\pm}_{\text{II}}(\vec{R})&=&\sum_{l=-\infty}^{\infty} d^{\pm
1}_{l}\frac{1}{2}\left(\chi^{+}_{l,\uparrow}(\vec{R},E-V_{0})+\chi^{-}_{l,\uparrow}(\vec{R},E-V_{0})\right)\nonumber\\
&+&d^{\pm
2}_{l}\frac{1}{2}\left(\chi^{+}_{l,\downarrow}(\vec{R},E-V_{0})+\chi^{-}_{l,\downarrow}(\vec{R},E-V_{0})\right)
\label{eq:waveinside}
\end{eqnarray}
Note that Eq.~(\ref{eq:waveinside}) contains both incoming and
outgoing states [as given in Eq.~(\ref{eq:polstates})] in order to
remove the $Y_{l}$ terms in $\chi^{\pm}_{l,\uparrow(\downarrow)}$,
which are singular at $\vec{R}=\vec{r}_{k}$.

From the continuity equations of the Schrodinger equations,
$\Psi^{\pm}_{\text{I}}(\vec{R})$ and
$\Psi^{\pm}_{\text{II}}(\vec{R})$ must satisfy the following
conditions for all $\vec{R}$ such that $|\vec{R}-\vec{r}_{k}|=a$:
\begin{eqnarray}
\Psi^{\pm}_{\text{I}}(|\vec{R}-\vec{r}_{k}|=a)=\Psi^{\pm}_{\text{II}}(|\vec{R}-\vec{r}_{k}|=a)
\label{eq:Cont1}
\end{eqnarray}
\begin{eqnarray}
\frac{\hbar^{2}}{2m^{*}}\frac{\partial\Psi^{\pm}_{\text{I}}(\vec{R})}{\partial
r}|_{|\vec{R}-\vec{r}_{k}|=a}&+&\left(\frac{\hbar^{2}}{2m^{*}a}-\text{i}\alpha\left(\begin{array}{cc}0&\exp(\text{i}\theta^{\vec{R}}_{\vec{r}_{k}})\\\exp(-\text{i}\theta^{\vec{R}}_{\vec{r}_{k}})&0\end{array}\right)\right)\Psi^{\pm}_{\text{I}}(|\vec{R}-\vec{r}_{k}|=a)
=\nonumber\\
\frac{\hbar^{2}}{2m^{*}}\frac{\partial\Psi^{\pm}_{\text{II}}(\vec{R})}{\partial
r}|_{|\vec{R}-\vec{r}_{k}|=a}&+&\left(\frac{\hbar^{2}}{2m^{*}a}-\text{i}\alpha_{\text{II}}\left(\begin{array}{cc}0&\exp(\text{i}\theta^{\vec{R}}_{\vec{r}_{k}})\\\exp(-\text{i}\theta^{\vec{R}}_{\vec{r}_{k}})&0\end{array}\right)\right)\Psi^{\pm}_{\text{II}}(|\vec{R}-\vec{r}_{k}|=a)
\nonumber\\\label{eq:Cont2}
\end{eqnarray}
In the following discussion, $\alpha_{II}=\alpha$, i.e., the
spin-orbit coupling strength is the same inside and outside the
well.

The solutions for the various coefficients $d^{\pm}_{l}$ and
$f^{\pm}_{l}$ are given in Appendix A for a cylindrical
well/barrier. In the following, we are interested in studying the
wave function away from the scatterer, so the relevant coefficients
are $f^{\pm 1}_{l}$ and $f^{\pm 2}_{l}$, which can be written for
convenience as
\begin{eqnarray}
f^{\pm
1}_{l}&=&2\text{i}^{l}\exp(\text{i}\vec{k}_{\pm}\cdot\vec{r}_{k})\exp(-\text{i}l\theta_{0})\frac{\widetilde{f}^{\pm
1}_{l}}{\sqrt{k_{1}}}\nonumber\\
 f^{\pm
2}_{l}&=&2\text{i}^{l}\exp(\text{i}\vec{k}_{\pm}\cdot\vec{r}_{k})\exp(-\text{i}l\theta_{0})\frac{\widetilde{f}^{\pm
2}_{l}}{\sqrt{k_{2}}}
\end{eqnarray}
The coefficients, $\widetilde{f}^{\pm 1}_{l}$ and
$\widetilde{f}^{\pm 2}_{l}$, depend upon the energy and the form of
the potential but do not depend upon the initial direction of the
incident momentum vector, $\theta_{0}$ (different potentials will
generate a different dependence of the coefficients upon
$\theta_{0}$, $E$, etc.). Thus the scattered wave function,
$\Psi^{\pm}_{S}(\vec{R})$ can be written as
\begin{eqnarray}
\Psi^{\pm}_{S}(\vec{R})&=&\sum_{l=-\infty}^{\infty}\text{i}^{l}\exp\left(\text{i}l\left[\theta^{\vec{R}}_{\vec{r}_{k}}-\theta_{0}\right]\right)\left(\widetilde{T}_{k,l}\Phi^{\pm}_{\text{in}}(\vec{R})\right)_{\vec{R}=\vec{r}_{k}}
\label{eq:talmost1}
\end{eqnarray}
where
\begin{eqnarray}
\widetilde{T}_{k,l}&=&\left(\widetilde{f}^{+1}_{l}\chi^{+}_{l,\uparrow}(\vec{R}-\vec{r}_{k},E)\langle
+(\theta_{0})|+\widetilde{f}^{-1}_{l}\chi^{+}_{l,\uparrow}(\vec{R}-\vec{r}_{k},E)\langle
-(\theta_{0})|\right)\nonumber\\
&+&\left(\widetilde{f}^{+2}_{l}\chi^{+}_{l,\downarrow}(\vec{R}-\vec{r}_{k},E)\langle
+(\theta_{0})|+\widetilde{f}^{-2}_{l}\chi^{+}_{l,\downarrow}(\vec{R}-\vec{r}_{k},E)\langle
-(\theta_{0})|\right)\nonumber\\
&=&\frac{1}{2}\left(\begin{array}{cc}H_{l}(k_{1}r_{\vec{R},\vec{r}_{k}})A^{1}_{l}&\exp(\text{i}\theta_{0})H_{l}(k_{1}r_{\vec{R},\vec{r}_{k}})B^{1}_{l}\\
-\exp(-\text{i}\theta^{\vec{R}}_{\vec{r}_{k}})\text{i}H_{l-1}(k_{1}r_{\vec{R},\vec{r}_{k}})A^{1}_{l}&-\exp(\text{i}(\theta_{0}-\theta^{\vec{R}}_{\vec{r}_{k}}))\text{i}H_{l-1}(k_{1}r_{\vec{R},\vec{r}_{k}})B^{1}_{l}\end{array}\right)\nonumber\\
&+&\frac{1}{2}\left(\begin{array}{cc}H_{l}(k_{2}r_{\vec{R},\vec{r}_{k}})A^{2}_{l}&\exp(\text{i}\theta_{0})H_{l}(k_{2}r_{\vec{R},\vec{r}_{k}})B^{2}_{l}\\
\exp(-\text{i}\theta^{\vec{R}}_{\vec{r}_{k}})\text{i}H_{l-1}(k_{2}r_{\vec{R},\vec{r}_{k}})A^{2}_{l}&\exp(\text{i}(\theta_{0}-\theta^{\vec{R}}_{\vec{r}_{k}}))\text{i}H_{l-1}(k_{2}r_{\vec{R},\vec{r}_{k}})B^{2}_{l}\end{array}\right)\nonumber\\
\label{eq:Talmost}
\end{eqnarray}
where
\begin{eqnarray}
\widetilde{A}^{j}_{l}&=&\widetilde{f}^{+j}_{l}+\widetilde{f}^{-j}_{l}\nonumber\\
\widetilde{B}^{j}_{l}&=&\widetilde{f}^{+j}_{l}-\widetilde{f}^{-j}_{l}\end{eqnarray}
where $H_{l}\equiv H^{+}_{l}$ in Eq.~(\ref{eq:Talmost}) (the $+$
sign will be implicitly assumed for the Hankel function for the rest
of this paper).

The operators $\widetilde{T}_{k,l}$ in Eq.~(\ref{eq:talmost1}) and
Eq.~(\ref{eq:Talmost}) appear to generate the l$^{th}$ partial wave
from the incident wave function, $\Phi_{\text{in}}$, evaluated at
the site of the scatterer. The only problem with this interpretation
are the various factors of $\exp(\text{i}l\theta_{0})$ occurring in
Eqs. (\ref{eq:talmost1})-(\ref{eq:Talmost}).   Since different
incident waves will possess or be a superposition of different
incident momentum directions (i.e., $\theta_{0}$ in Figure
\ref{fig:figure1}), an additional operator needs to be constructed
which generates the various factors of $\exp(\text{i}l\theta_{0})$
from the incident wave with energy $E$. The operator,
$\widehat{D}_{l}$, can be constructed such that for any given state
of the form
$|\Phi_{\text{in}}\rangle=c_{1}|k_{1}(\theta_{0}),+(\theta_{0})\rangle+c_{2}|k_{2}(\theta_{0}),-(\theta_{0})\rangle$,
\begin{eqnarray}
\widehat{D}_{l}\Phi_{\text{in}}(\vec{R})&=&\exp(\text{i}l\theta_{0})\Phi_{\text{in}}(\vec{R})
\end{eqnarray}
The operator, $\widehat{D}_{l}$ which satisfies the above equation
is given by
\begin{eqnarray}
\widehat{D}_{l}&=&\left(\begin{array}{cc}a_{l}\widehat{P}_{l}&b_{l}\widehat{P}_{l+1}\\
c_{l}\widehat{P}_{l-1}&d_{l}\widehat{P}_{l}\end{array}\right)
\end{eqnarray}
where \begin{eqnarray}
\widehat{P}_{l}&=&\left(\frac{l}{|l|}\frac{\partial}{\partial(\vec{R}\cdot\widehat{X})}-\text{i}\frac{\partial}{\partial(\vec{R}\cdot\widehat{Y})}\right)^{|l|}
\end{eqnarray}
 with $\widehat{P}_{0}=1$.  The construction
and full expression for $\widehat{D}_{l}$ is given in the Appendix
B.

The operator which generates the scattered wave from the wave
incident upon scatterer $k$, $\widehat{T}_{k}$, can finally be
written as
\begin{eqnarray}
\widehat{T}_{k}&=&\sum_{l=-\infty}^{\infty}\text{i}^{l}\exp\left(-\text{i}l\theta^{\vec{R}}_{\vec{r}_{k}}\right)\widehat{G}^{k}_{l}\widehat{P}_{l}
\label{eq:Toperator}
\end{eqnarray}
where
\begin{eqnarray}
\widehat{G}^{k}_{l}&=&\frac{1}{2}\left(\begin{array}{cc}H_{l}(k_{1}r_{\vec{R},\vec{r}_{k}})
&-\text{i}\exp(\text{i}\theta^{\vec{R}}_{\vec{r}_{k}})H_{l-1}(k_{1}r_{\vec{R},\vec{r}_{k}})\\
\text{i}\exp(-\text{i}\theta^{\vec{R}}_{\vec{r}_{k}})H_{l+1}(k_{1}r_{\vec{R},\vec{r}_{k}})&
H_{l}(k_{1}r_{\vec{R},\vec{r}_{k}})\end{array}\right)\left(\begin{array}{cc}
\widehat{t}^{11}_{k,l}&0\\0&\widehat{t}^{12}_{k,l}\end{array}\right)\nonumber\\
&+&\frac{1}{2}\left(\begin{array}{cc}H_{l}(k_{2}r_{\vec{R},\vec{r}_{k}})&\text{i}\exp(\text{i}\theta)H_{l-1}(k_{2}r_{\vec{R},\vec{r}_{k}})\\
-\text{i}\exp(-\text{i}\theta)H_{l+1}(k_{2}r_{\vec{R},\vec{r}_{k}})&H_{l}(k_{2}r_{\vec{R},\vec{r}_{k}})\end{array}\right)
\left(\begin{array}{cc}\widehat{t}^{21}_{k,l}&0\\0&\widehat{t}^{22}_{k,l}\end{array}\right)\end{eqnarray}
where $t^{11}_{k,l}=A^{k,1}_{-l}a_{l}+B^{k,1}_{-l}c_{1+l}$,
$t^{l2}_{k,l}=A^{k,1}_{1-l}b_{l-1}+B^{k,1}_{1-l}d_{l}$,
$t^{21}_{k,l}=A^{k,2}_{-l}a_{l}+B^{k,2}_{-l}c_{1+l}$, and
$t^{22}_{k,l}=-A^{k,2}_{1-l}b_{l-1}-B^{k,2}_{1-l}d_{l}$.  From the
values for the various $f^{k,\pm 1}_{l}$ and $f^{k,\pm 2}_{l}$
calculated in Appendix A for a cylindrically symmetric barrier/well,
it can be shown that $t^{11}_{k,l}=t^{12}_{k,-l}$ and
$t^{21}_{k,l}=t^{22}_{k,-l}$.  Note that the form of
$\widehat{T}_{k}$ is the same for all cylindrically symmetric
scatterers;  the values of the scattering amplitudes,
$\widetilde{f}^{\pm j}_{l}$, depend upon the actual potential used
for the cylindrically symmetric scatterer.
\section{Multiple Scattering theory}
For $N$ isolated scatterers, the overall wave function at
$\vec{R}$ can be written as
\begin{eqnarray}
\Psi(\vec{R})&=&\Phi_{\text{in}}(\vec{R})+\sum_{k=1}^{N}(\widehat{T}_{k}\Psi(\vec{R}))_{\vec{R}=\vec{r}_{k}}
\label{eq:multiscateq}
\end{eqnarray}
Equation (\ref{eq:multiscateq}) indicates that if the value of
$\Psi(\vec{R})$ and its derivatives [due to the $\widehat{P}_{l}$
dependence of $\widehat{T}_{k}$ in Eq.~(\ref{eq:Toperator})] at each
scatterer is known, the entire wave function $\Psi(\vec{R})$ is
completely determined. In principle, the values of $\Psi(\vec{R})$
and its derivatives at each scatterer can be found using
Eq.~(\ref{eq:multiscateq}).  In practice, it is only practical to
calculate the first few derivatives of $\Psi(\vec{R})$ at each
scatterer. When the size of the scatterer (or in general, the
scattering length) is much smaller than the wavelengths, i.e.,
$k_{1}a\ll 1$ and $k_{2}a\ll 1$, the only significant contribution
to $\widehat{T}_{k}$ comes from the $l=0$ term in
Eq.~(\ref{eq:Toperator}). This is analogous to the heavily studied
``s"-wave scattering models, and in the following discussion, only
the $l=0$ term in Eq.~(\ref{eq:Toperator}) will be considered.

\section{Low energy scattering limit}
In the limit $k_{1(2)}a\ll 1$, Eq.~(\ref{eq:multiscateq}) can be
approximated as
\begin{eqnarray}
\Psi(\vec{R})&=&\Phi_{\text{in}}(\vec{R})+\sum_{k=1}^{N}G^{k}_{0}(\vec{R})\Psi(\vec{r}_{k})
\label{eq:swavey}
\end{eqnarray}
where $G^{k}_{0}$ can be written as
\begin{eqnarray}
G^{k}_{0}(\vec{R})&=&\frac{1}{2}\left(\begin{array}{cc}H_{0}(k_{1}r_{\vec{R},\vec{r}_{k}})&\text{i}\exp(\text{i}\theta^{\vec{R}}_{\vec{r}_{k}})H_{1}(k_{1}r_{\vec{R},\vec{r}_{k}})\\
\text{i}\exp(-\text{i}\theta^{\vec{R}}_{\vec{r}_{k}})H_{1}(k_{1}r_{\vec{R},\vec{r}_{k}})&H_{0}(k_{1}r_{\vec{R},\vec{r}_{k}})\end{array}\right){t}^{1}_{k,0}\nonumber\\
&+&\frac{1}{2}\left(\begin{array}{cc}H_{0}(k_{2}r_{\vec{R},\vec{r}_{k}})&-\text{i}\exp(\text{i}\theta^{\vec{R}}_{\vec{r}_{k}})H_{1}(k_{2}r_{\vec{R},\vec{r}_{k}})\\
-\text{i}\exp(-\text{i}\theta^{\vec{R}}_{\vec{r}_{k}})H_{1}(k_{2}r_{\vec{R},\vec{r}_{k}})&H_{0}(k_{2}r_{\vec{R},\vec{r}_{k}})\end{array}\right)t^{2}_{k,0}
\label{eq:GGG}
\end{eqnarray}
It should be noted that Eq.~(\ref{eq:swavey}) and Eq.~(\ref{eq:GGG})
are similar to the Lippmann-Schwinger equation for a potential,
$V(\vec{r})$ comprised of $N$ delta-functions:
$V_{\delta}(\vec{r})=\sum_{k}V_{k}\delta(\vec{r}-\vec{r}_{k})$,
which has been used before in previous studies of the spin dynamics
in the presence of spin-orbit
coupling\cite{Mishchenko04,Inoue03,Schliemann03}.
 From the Lippmann-Schwinger equation, the wave function in the presence of $V_{\delta}(\vec{r})$ is given by
 \begin{eqnarray}
 \Psi(\vec{R})&=&\Phi_{\text{in}}(\vec{R})+\int \widehat{G}_{+}(\vec{R},\vec{r},E)V_{\delta}(\vec{r})\Psi(\vec{r})\text{d}^{3}r\nonumber\\
&=&\Phi_{\text{in}}(\vec{R})+\sum_{k=1}^{N}V_{k}\widehat{G}_{+}(\vec{R},\vec{r}_{k},E)\Psi(\vec{r}_{k})
\end{eqnarray}
where $\widehat{G}_{+}(\vec{R},\vec{r}_{k},E)$ is the Green's
function in the presence of Rashba spin-orbit coupling, which is
given by Eq.~(\ref{eq:GF}) in Appendix C and is similar in form to
$G^{k}_{0}(\vec{R})$.  The form of $G^{k}_{0}(\vec{R})$ would be
identical to $\widehat{G}_{+}(\vec{R},\vec{r}_{k})$ if
\begin{eqnarray}
\frac{t_{k,0}^{1}}{t_{k,0}^{2}}&=&\frac{k_{1}}{k_{2}}
\label{eq:sameo}
\end{eqnarray}
In general, Eq.~(\ref{eq:sameo}) is not satisfied, although for
$\overline{k}a\ll 1$ Eq.~(\ref{eq:sameo}) is approximately correct.
The difference $\widehat{G}_{+}(\vec{R},\vec{r}_{k},E)$, and
$\widehat{G}_{0}^{k}(\vec{R})$ can be understood as follows:
$\widehat{G}_{+}(\vec{R},\vec{r}_{k},E)$ propagates the scattered
wave function from the a delta-function potential, whereas
$\widehat{G}_{0}^{k}(\vec{R})$ propagates the scattered wave
function from the finite-sized potential, $V_{k}(\vec{r})$ [Eq.
(\ref{eq:pot})], which, in the delta function limit ($a\rightarrow
0$, $V_{0}\rightarrow\pm \infty$, $\pi V_{0}a^{2}=\pm V_{k}$),
doesn't scatter (i.e., the scattering coefficients, $t^{1}_{k,0}$
and $t^{2}_{k,0}$, vanish).

Far away from the scatterers ($k_{1(2)}r_{\vec{R},\vec{r}_{k}}\gg
1$), Eq.~(\ref{eq:GGG}) can be written as
\begin{eqnarray}
G^{k}_{0}(\vec{R})&=&\sqrt{\frac{2}{\pi
r_{\vec{R},\vec{r}_{k}}}}\exp\left(\text{i}\left[\overline{k}r_{\vec{R},\vec{r}_{k}}-\frac{\pi}{4}+\phi_{k}\right]\right)\left(\overline{t}_{k}-\text{i}\overline{\delta
t}_{k}\widehat{U}\left(\frac{\pi}{2},\theta^{\vec{R}}_{\vec{r}_{k}}\right)\right)\widehat{U}\left(k_{\alpha}r_{\vec{R},\vec{r}_{k}}+\delta\phi_{k},\theta^{\vec{R}}_{\vec{r}_{k}}\right)\nonumber\\
\end{eqnarray}
where \begin{eqnarray}
\overline{k}&=&\frac{k_{1}+k_{2}}{2}=\sqrt{\left(\frac{m^{*}\alpha}{\hbar^{2}}\right)^{2}+\frac{2m^{*}}{\hbar^{2}}E}\nonumber\\
k_{\alpha}&=&\frac{k_{1}-k_{2}}{2}=\frac{\alpha m^{*}}{\hbar^{2}}\nonumber\\
\overline{t}_{k}&=&\frac{|\widehat{t}^{1}_{k,0}|\sqrt{k_{2}}+|\widehat{t}^{2}_{k,0}|\sqrt{k_{1}}}{2\sqrt{k_{1}k_{2}}}\nonumber\\
\overline{\delta
t_{k}}&=&\frac{|\widehat{t}^{1}_{k,0}|\sqrt{k_{2}}-|\widehat{t}^{2}_{k,0}|\sqrt{k_{1}}}{2\sqrt{k_{1}k_{2}}}\nonumber\\
\exp(\text{i}\delta\phi_{k})&=&\sqrt{\frac{\widehat{t}^{1}_{k,0}\left(\widehat{t}^{2}_{k,0}\right)^{*}}{|\widehat{t}^{1}_{k,0}\widehat{t}^{2}_{k,0}|}}\nonumber\\
\exp(\text{i}\phi_{k})&=&\sqrt{\frac{\widehat{t}^{1}_{k,0}\widehat{t}^{2}_{k,0}}{|\widehat{t}^{1}_{k,0}\widehat{t}^{2}_{k,0}|}}\nonumber\\
\label{eq:params}
\end{eqnarray}
and $\widehat{U}(\theta,\phi)$ is a rotation operator given
by\begin{eqnarray}
\widehat{U}(\theta,\phi)&=&\exp\left(\text{i}\frac{\phi}{2}\widehat{\sigma}_{Z}\right)\exp\left(\text{i}\theta\widehat{\sigma}_{X}\right)\exp\left(-\text{i}\frac{\phi}{2}\widehat{\sigma}_{Z}\right)
\end{eqnarray}
Eq.~(\ref{eq:GGG}) contains a dynamical factor which depends upon
the distance from scatterer k multiplied by a sum of two rotation
operators. In the presence of spin-orbit coupling, the low energy
limit scattered wave functions possess an ``s"-wave character and
also a ``p"-wave character due to the
$\theta^{\vec{R}}_{\vec{r}_{k}}$ dependence in Eq.~(\ref{eq:GGG}),
which vanishes in the limit as $\alpha\rightarrow 0$.

In the calculations to be performed, $\delta
t_{k}/\overline{t}_{k}\approx 0.05$ and $\delta\phi_{k}\ll 1$. In
this case, Eq.~(\ref{eq:GGG}) can be approximately written as [for
$k_{1(2)}|\vec{R}-\vec{r}_{k}|\gg 1$]:
\begin{eqnarray}
G^{k}_{0}(\vec{R})&=&\overline{t}_{k}\sqrt{\frac{2}{\pi
r_{\vec{R},\vec{r}_{k}}}}\exp\left(\text{i}\left[\overline{k}r_{\vec{R},\vec{r}_{k}}-\frac{\pi}{4}+\phi_{k}\right]\right)\widehat{U}\left(k_{\alpha}r_{\vec{R},\vec{r}_{k}},\theta^{\vec{R}}_{\vec{r}_{k}}\right)
\label{eq:GGGapprox}
\end{eqnarray}
which now contains a dynamical, distance-dependent factor times a
single rotation operator,
$\widehat{U}(k_{\alpha}r_{\vec{R},\vec{r}_{k}},\theta^{\vec{R}}_{\vec{r}_{k}})$,
which corresponds to a rotation by an angle
$k_{\alpha}r_{\vec{R},\vec{r}_{k}}$ about the spin-orbit field for
propagation along the direction
$\frac{\vec{R}-\vec{r}_{k}}{|\vec{R}-\vec{r}_{k}|}$. In this limit,
the total wave function in the presence of $N$ scatterers is given
by
\begin{eqnarray}
\Psi(\vec{R})&=&\Phi(\vec{R})+\sum_{j}\overline{t}_{j}\sqrt{\frac{2}{\pi
r_{\vec{R},\vec{r}_{j}}}}\exp\left(\text{i}\left[\overline{k}r_{\vec{R},\vec{r}_{j}}-\frac{\pi}{4}+\phi_{j}\right]\right)\widehat{U}(k_{\alpha}r_{\vec{R},\vec{r}_{j}},\theta^{\vec{R}}_{\vec{r}_{j}})\Psi(\vec{r}_{j})
\label{eq:waveo}
\end{eqnarray}

From Eq.~(\ref{eq:waveo}), knowing the value of the wave function at
each scatterer $k$, $\Psi(\vec{r}_{k})$, completely determines the
total wave function $\Psi(\vec{R})$. The various values of
$\Psi(\vec{r}_{k})$ can be found by setting $\vec{R}=\vec{r}_{k}$ in
Eq.~(\ref{eq:swavey}) for each scatterer $k$.  This provides a
system of $2N$ linear equations to solve for the various
$\Psi(\vec{r}_{k})$. The resulting system of equations can be
expressed in matrix form as
\begin{eqnarray}
\widehat{M}\widehat{\Psi}&=&\widehat{\phi}
\end{eqnarray}
where $\widehat{\Psi}$ and $\widehat{\phi}$ are $2N$ by 1 matrices
where $\widehat{\Psi}(2k-1)=\Psi^{\uparrow}(\vec{r}_{k})$,
$\widehat{\Psi}(2k)=\Psi^{\downarrow}(\vec{r}_{k})$,
 $\widehat{\phi}(2k-1)=\phi^{\uparrow}(\vec{r}_{k})$,
 $ \widehat{\phi}(2k)=\phi^{\downarrow}(\vec{r}_{k})$, and
$\widehat{M}$ is a $2N$ by $2N$ matrix where $\widehat{M}(m,m)=1$
for $m=1$ to $m=2N$, and for $k,j\in[1,N]$ and $k\neq j$,
\begin{eqnarray}
\widehat{M}(2k-1,2j-1)&=&-\left(G^{j}_{0}(\vec{r}_{k})\right)_{1,1}\nonumber\\
\widehat{M}(2k,2j)&=&-\left(G^{j}_{0}(\vec{r}_{k})\right)_{2,2}\nonumber\\
\widehat{M}(2k-1,2j)&=&-\left(G^{j}_{0}(\vec{r}_{k})\right)_{1,2}\nonumber\\
\widehat{M}(2k,2j-1)&=&-\left(G^{j}_{0}(\vec{r}_{k})\right)_{2,1}\nonumber\\
\end{eqnarray}
where $G^{j}_{0}(\vec{r}_{k})$ is given in Eq.~(\ref{eq:GGG}). Once
$\widehat{M}$ is specified, $\widehat{\Psi}$ can be found by
inverting $\widehat{M}$ as follows:
\begin{eqnarray}
\widehat{\Psi}&=&\widehat{M}^{-1}\widehat{\phi}
\label{eq:finalcdown}
\end{eqnarray}
\section{Applications to flux measurements in two-dimensional electron gases
in the presence of Rashba spin-orbit coupling} Recent
experiments\cite{Topinka00,Topinka01,Topinka03} have imaged electron
flow in a 2DEG by monitoring the changes in conductance through a
single quantum point contact (QPC) as a moveable SPM tip is scanned
above the surface of a heterostructure.  By applying a negative
voltage between the SPM tip and the 2DEG,  a potential that
electrons in the 2DEG can scatter from is generated.
 As the SPM tip is scanned over the surface of the heterostructure,
 the backscattered current into the QPC is monitored, which is
proportional to the the electron flow in the probed region in the
absence of the tip. In this manner, an image of the electron flow is
produced. On top of the electron flow pattern, interference fringes
can also be observed in these experiments, which are due to the
interference between the single scattering trajectory from the SPM
tip with the single scattering trajectories from random impurities
within the sample. Recently, additional interference effects for
electrons emitted from a single QPC were also reported for an SPM
tip in the presence of a fixed reflector gate\cite{LeRoy05}, where
interference fringes result not only from the interference between
the single scattering trajectories of the reflector gate and the SPM
tip but also from the interference of the double scattering
trajectories involving the reflector gate and the single scattering
trajectory from the SPM tip.

 The samples used in the above
experimental studies were GaAs/AlGaAs heterostructures, which have
very low spin-orbit coupling\cite{Miller03}
($\alpha=3\times10^{-13}$ eVm), and the experimental results were
well described by quantum simulations/calculations in the absence of
spin-orbit
coupling\cite{Topinka00,Topinka01,LeRoy05,shawpap,Shawthesis,Heller05}.
 However, other samples can
possess considerably larger spin-orbit coupling, \cite{Grundler00}
such as InAs, which can have $\alpha=4\times10^{-11}$ eVm.  The
question therefore arises as to what effects or signatures of
spin-orbit coupling exist in electron flow imaging experiments using
a moveable SPM tip.
\begin{figure}[tbp]
\centering \includegraphics[width=10cm]{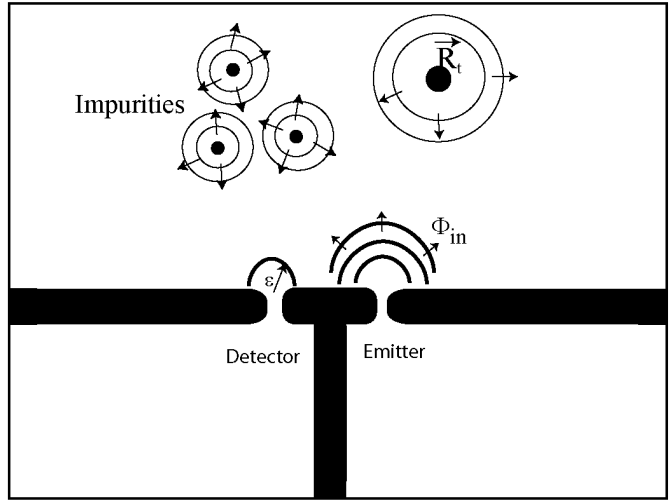}
\caption{Electrons injected by the emitter QPC (represented by
$\Phi_{\text{in}}$) are backscattered due to random impurities in
the 2DEG along with a moveable ``scatterer" generated by an SPM
above the 2DEG surface. Measurements of the backscattered flux into
the detector QPC as a function of the SPM tip position,
$\vec{R}_{t}$, can be used to image the electron flow from the
emitter.} \label{fig:figure2}
\end{figure}

Figure \ref{fig:figure2} shows the setup under consideration.  A
point source is used to inject electrons into the 2DEG (analogous to
a QPC). The injected electrons are backscattered by random
impurities present in the sample and by the potential generated by
the SPM tip placed above the surface of the 2DEG. The backscattered
current into the detection QPC is then measured as a function of the
SPM tip position, $\vec{R}_{t}$.  In the setup shown in Figure
\ref{fig:figure2}, the possibility that the detection QPC can be
separate from the emitter QPC is allowed. Such experimental
geometries have been used in magnetic focussing experiments in the
past\cite{Houten89}.

The injected electrons from the emitter are taken to be an
unpolarized beam comprised of an equal mixture of the cylindrical
wave-like states $\Phi_{1}(\vec{R})$ and $\Phi_{2}(\vec{R})$, which
are given by
\begin{eqnarray}
\label{eq:phi1na}
\Phi_{1}(\vec{R})&=&\exp\left(\text{i}\frac{\pi}{4}\right)\left(\chi_{0,\uparrow}
(\vec{R})+\chi_{0,\downarrow}(\vec{R})\right)\nonumber\\
\Phi_{2}(\vec{R})&=&
\text{i}\exp\left(\text{i}\frac{\pi}{4}\right)\left(
\chi_{1,\uparrow}(\vec{R})-\chi_{1,\downarrow}(\vec{R})\right)
\label{eq:phi2na}
\end{eqnarray}
Each state represents current of $\hbar\overline{k}/m^{*}$ being
injected into the emitter, as shown in Figure \ref{fig:figure2}. Far
away from the source, the states can be approximated as
\begin{eqnarray}
\label{eq:phi1}
\Phi_{1}(\vec{R})&=&\frac{1}{\sqrt{\pi|\vec{R}|}}\exp\left(\text{i}\overline{k}|\vec{R}|\right)\widehat{U}(k_{\alpha}|\vec{R}|,\theta_{e})\binom{1}{0}\\
\Phi_{2}(\vec{R})&=&\frac{1}{\sqrt{\pi|\vec{R}|}}\exp\left(\text{i}\overline{k}|\vec{R}|\right)\widehat{U}(k_{\alpha}|\vec{R}|,\theta_{e})\binom{0}{1}\label{eq:phi2}
\end{eqnarray}
where $\theta_{e}$ is the given with respect to the emitter.  Note
that when $\alpha\rightarrow 0$, the states $\Phi_{1}$ and
$\Phi_{2}$ correspond to a spin up and spin down electron being
injected from the point source.

The net current injected into the detector is given by the following
formula:\begin{eqnarray}
\mu&=&\epsilon\int^{\frac{\pi}{2}}_{-\frac{\pi}{2}}\text{d}\theta\left(J_{X}(\vec{\epsilon})\sin(\theta)+J_{Y}(\vec{\epsilon})\cos(\theta)\right)
\label{eq:current}
\end{eqnarray}
where
\begin{eqnarray}
J_{X}(\vec{\epsilon})&=&\frac{\hbar}{m^{*}}Im\left[\Psi^{\dagger}(\vec{R})\nabla_{X}\Psi(\vec{R})\right]_{\vec{R}=\vec{\epsilon}}+\frac{\alpha}{\hbar}\Psi^{\dagger}(\vec{\epsilon})\widehat{\sigma}_{Y}\Psi(\vec{\epsilon})\nonumber\\
J_{Y}(\vec{\epsilon})&=&\frac{\hbar}{m^{*}}Im\left[\Psi^{\dagger}(\vec{R})\nabla_{Y}\Psi(\vec{R})\right]_{\vec{R}=\vec{\epsilon}}-\frac{\alpha}{\hbar}\Psi^{\dagger}(\vec{\epsilon})\widehat{\sigma}_{X}\Psi(\vec{\epsilon})
\end{eqnarray}
and
$\vec{\epsilon}=\epsilon(\cos(\theta)\widehat{Y}+\sin(\theta)\widehat{X})$.
In the actual experiment/simulation, the current injected into the
detector is measured as a function of tip position several microns
away from the detector. As will be discussed later, the interference
between the incident waves and the scattered waves can be neglected
in Eq.~(\ref{eq:current}) due to thermal averaging, so only the
scattered wave function, $\Psi_{S}$, needs to be considered.
Phase-coherent transport is assumed in the application of Eq.
(\ref{eq:current}).  Additionally, since the width of the detector
is taken to be negligible, only the current operator of $\Psi_{S}$
evaluated at the detector needs to taken into account in
Eq.~(\ref{eq:current}). The injected current into the detector
becomes
\begin{eqnarray} \mu&=&2\epsilon\left(\frac{\hbar}{m^{*}}
Im\left[\Psi^{\dagger}_{\text{S}}(\vec{R}_{d})\nabla_{Y}
\Psi_{\text{S}}(\vec{R}_{d})\right]-\frac{\alpha}{\hbar}\Psi^{\dagger}_{\text{S}}(\vec{R}_{d})\widehat{\sigma}_{X}\Psi_{\text{S}}(\vec{R}_{d})\right)
\end{eqnarray}
Using the form of the wave function in Eq.~(\ref{eq:waveo})
evaluated at the site of the detector, the injected current can then
be written as a function of tip position and energy as
\begin{eqnarray}
\mu(\vec{R}_{t},E)&=&\frac{4\epsilon}{\pi}\sum_{k}\frac{\overline{t}_{k}^{2}}{r_{k,d}}\cos(\theta_{k}^{d})\frac{\hbar\overline{k}}{m^{*}}\Psi^{\dagger}(\vec{r}_{k})\Psi(\vec{r}_{k})\nonumber\\
&-&\frac{\overline{t}^{2}_{k}}{r_{k,d}}\frac{\alpha}{\hbar}\Psi^{\dagger}(\vec{r}_{k})\widehat{U}^{\dagger}(k_{\alpha}r_{k,d},\theta^{d}_{k})\left[\frac{\sin(2\theta_{k}^{d})}{2}\widehat{\sigma}_{Y}+\sin^{2}(\theta_{k}^{d})\widehat{\sigma}_{X}\right]\widehat{U}(k_{\alpha}r_{k,d},\theta^{d}_{k})\Psi(\vec{r}_{k})\nonumber\\
&+&\frac{4\epsilon}{\pi}\frac{\hbar\overline{k}}{m^{*}}\sum_{j<k}\overline{t}_{k}\overline{t}_{j}\left(\frac{\cos(\theta_{k}^{d})+\cos(\theta_{j}^{d})}{2\sqrt{r_{k,d}r_{j,d}}}\right)\times\nonumber\\
&&\left(\Psi^{\dagger}(\vec{r}_{k})\widehat{U}^{\dagger}(k_{\alpha}r_{k,d},\theta_{k}^{d})\widehat{U}(k_{\alpha}r_{j,d},\theta_{j}^{d})\Psi(\vec{r}_{j})\exp(\text{i}[\overline{k}(r_{j,d}-r_{k,d})+\phi_{j}-\phi_{k}])+h.c.\right)\nonumber\\
&+&\frac{4\epsilon}{\pi}\frac{\alpha}{\hbar}\sum_{j<k}\frac{\overline{t}_{k}\overline{t}_{j}}{\sqrt{r_{j,d}r_{k,d}}}\left(\frac{(\cos(\theta_{j}^{d}))^{2}+(\cos(\theta_{k}^{d}))^{2}}{2}-1\right)\times\nonumber\\
&&\left[\Psi^{\dagger}(\vec{r}_{k})\widehat{U}^{\dagger}(k_{\alpha}r_{k,d},\theta_{k}^{d})\widehat{\sigma}_{X}\widehat{U}(k_{\alpha}r_{j,d},\theta_{j}^{d})\Psi(\vec{r}_{j})\exp(\text{i}\overline{k}(r_{j,d}-r_{k,d})+\phi_{j}-\phi_{k})+h.c.\right]\nonumber\\
&-&\frac{4\epsilon}{\pi}\frac{\alpha}{\hbar}\sum_{j<k}\frac{\overline{t}_{k}\overline{t}_{j}}{\sqrt{r_{j,d}r_{k,d}}}\frac{\sin(\theta_{j}^{d})\cos(\theta_{j}^{d})+\sin(\theta_{k}^{d})\cos(\theta_{k}^{d})}{2}\times\nonumber\\
&&\left[\Psi^{\dagger}(\vec{r}_{k})\widehat{U}^{\dagger}(k_{\alpha}r_{k,d},\theta_{k}^{d})\widehat{\sigma}_{Y}\widehat{U}(k_{\alpha}r_{j,d},\theta_{j}^{d})\Psi(\vec{r}_{j})\exp(\text{i}\overline{k}(r_{j,d}-r_{k,d})+\phi_{j}-\phi_{k})+h.c.\right]\nonumber\\
\label{eq:gencond}
\end{eqnarray}
where the energy dependence of $\mu(\vec{R}_{t},E)$ comes in through
the energy dependence of both $\overline{k}$ and $\overline{t}_{k}$.
Since all experiments are done at nonzero temperatures, thermal
averaging of Eq.~(\ref{eq:gencond}) becomes necessary. Assuming the
injected current is a result of a small potential drop, $\delta V$,
over the emitter QPC and that the electrons on both sides of the
emitter QPC can be described as being free 2DEG, the energy
weighting function is simply related to the fermi-dirac distribution
function, $f(E)$, and is given by:
\begin{eqnarray}
-f^{'}(E)\text{d}E&=&\frac{\text{d}E}{k_{B}T}\frac{\exp\left(\frac{E-E_{F}}{k_{B}T}\right)}{\left(1+\exp\left(\frac{E-E_{F}}{k_{B}T}\right)\right)^{2}}
\label{eq:dis}
\end{eqnarray}
where $E_{F}$ is the Fermi energy.  All quantities calculated will
be thermally averaged using Eq.~(\ref{eq:dis}).  The thermally
averaged injected current is then given by
\begin{eqnarray}
\overline{\mu(\vec{R}_{t})}&=&-\int^{\infty}_{0}\mu(\vec{R}_{t},E)f^{'}(E)\text{d}E
\label{eq:thermie}
\end{eqnarray}

Since $\mu(\vec{R}_{t})$ contains interference terms that go like
$\overline{k}\exp(\text{i}\overline{k}r)$ (where $r$ is some length
related to the various distances in the system), the following
integral\cite{LandauSM1book} will be useful in evaluating
Eq.~(\ref{eq:thermie}):
\begin{eqnarray}
-\int^{\infty}_{0}\overline{k}
f^{'}(E)\exp(\text{i}\overline{k}r)\text{d}E&=&-2\text{i}\lambda_{T}\exp\left(\text{i}\overline{k}_{F}r\right)\sinh^{-1}(\lambda_{T}r)\left(1-\lambda_{T}r\text{coth}(\lambda_{T}r)\right)\nonumber\\
&+&\frac{\lambda_{T}}{\overline{k}_{F}}\exp\left(\text{i}\overline{k}_{F}r\right)\sinh^{-1}(\lambda_{T}r)\left(\overline{k}_{F}^{2}r-2\lambda_{T}\text{coth}(\lambda_{T}r)\right)\nonumber\\
&-&\frac{\lambda_{T}}{\overline{k}_{F}}\exp\left(\text{i}\overline{k}_{F}r\right)\sinh^{-1}(\lambda_{T}r)\lambda_{T}^{2}r\left(\text{coth}^{2}\left(\lambda_{T}r\right)+\sinh^{-2}\left(\lambda_{T}r\right)\right)\nonumber\\
&\equiv&\exp(\text{i}\overline{k}_{F}r)\sinh^{-1}(\lambda_{T}r)g(r,T,E_{F})
\end{eqnarray}
where $\lambda_{T}=\overline{k}_{F}\pi k_{B}T (2E_{F})^{-1}$.  For
large $r$, interference terms in $\mu(\vec{R}_{t})$ decay as
$r\exp(-\lambda_{T}r)$.  For $T=3 K$, $E_{F}=16$ meV,
$m^{*}=0.022m_{0}$ (where $m_{0}$ is the free electron mass),
$\lambda_{T}=2.35 \mu\text{m}^{-1}$, which allows one to neglect the
interference between the incoming wave and scattered wave when
calculating $\overline{\mu(\vec{R}_{t})}$ many microns away from the
QPC.
\subsection{The Single-scattering limit}
Before considering the case of multiple-scattering (which can only
be analytically solved for simple cases), it is useful to consider
the single-scattering case for an unpolarized beam (i.e., averaged
over the incident waves $\Phi_{1}$ [Eq.~(\ref{eq:phi1na})] and
$\Phi_{2}$ [Eq.~(\ref{eq:phi2na})]). In this case, the value of the
wave function at scatterer $k$ is given by
\begin{eqnarray}
\Psi_{1}(\vec{r}_{k})&=&\frac{1}{\sqrt{\pi
r_{e,k}}}\exp\left(\text{i}\overline{k}r_{e,k}\right)\widehat{U}(k_{\alpha}r_{e,k},\theta^{k}_{e})\binom{1}{0}
\label{eq:phia1}
\end{eqnarray}
for incident wave $\Phi_{1}$ [Eq.~(\ref{eq:phi1})] and
\begin{eqnarray}
\Psi_{2}(\vec{r}_{k})&=&\frac{1}{\sqrt{\pi
r_{e,k}}}\exp\left(\text{i}\overline{k}r_{e,k}\right)\widehat{U}(k_{\alpha}r_{e,k},\theta^{k}_{e})\binom{0}{1}
\label{eq:phia2}
\end{eqnarray}
for incident wave $\Phi_{2}$ [Eq.~(\ref{eq:phi2})].

First consider the case when the detector and the emitter are one
and the same. The spin averaged (i.e., averaged over incident waves
$\Phi_{1}$ and $\Phi_{2}$) and thermally averaged change in flux as
a function of tip position, $\overline{\Delta
\mu(\vec{R}_{t},E)}=\overline{\mu(\vec{R}_{t},E)}-\overline{\mu(\vec{R}_{t}=\infty,E)}$,
is given by:
\begin{eqnarray}
\overline{\Delta
\mu(\vec{R}_{t})}&\approx&\frac{2\epsilon}{\pi^{2}}\frac{\hbar\overline{k}_{F}}{m^{*}}\left(1+\frac{\pi^{2}}{12}\left(\frac{k_{B}T}{E_{F}}\right)^{2}\right)\overline{t}_{t}^{2}\frac{\cos(\theta_{t}^{d})}{(r_{tip,d})^2}\nonumber\\
&+&\frac{2\epsilon}{\pi^{2}}\frac{\hbar}{m^{*}}\sum_{k}\frac{\cos(\theta_{k}^d)+\cos(\theta_{t}^{d})}{r_{tip,d}r_{k,d}}\overline{t}_{k}\overline{t}_{t}
\exp(\text{i}2\overline{k}_{F}2(r_{tip,d}-r_{k,d})+\phi_{t}-\phi_{k})\nonumber\\
&\times&\sinh^{-1}\left[2\lambda_{T}(r_{tip,d}-r_{k,d})\right]g(2(r_{tip,d}-r_{k,d}),T,E_{F})+h.c.\nonumber\\
&=&\frac{2\epsilon}{\pi^{2}}\frac{\hbar\overline{k}_{F}}{m^{*}}\left(1+\frac{\pi^{2}}{12}\left(\frac{k_{B}T}{E_{F}}\right)^{2}\right)\overline{t}_{t}^{2}\frac{\cos(\theta_{t}^{d})}{(r_{tip,d})^2}+F(\vec{R}_{t})\exp\left(\text{i}2\overline{k}_{F}r_{tip,d}\right)+h.c.
\label{eq:fluxsame}\end{eqnarray} where $F(\vec{R}_{t})$ is some
nonoscillatory function which depends on the particular
configuration of scatterers, along with a some angular dependence of
the tip and a exponential damping factor depending upon the tip
position from the scatterers.  Spin-orbit effects are not seen in
Eq.~(\ref{eq:fluxsame}) due to the fact that for electrons moving
along effective one-dimensional trajectories, the amount of spin
rotation induced is simply proportional to the net distance the
electrons have traversed.  Since an electron which is directly
scattered back to the detector has effectively traveled no net
distance, the net spin rotation is zero. Another way to see this is
that
$\widehat{U}(k_{\alpha}r_{k,d},\theta_{d}^{k})\widehat{U}(k_{\alpha}r_{e,k},\theta_{e}^{k})=\widehat{1}$
when the emitter and detector are one and the same, since
$r_{k,d}=r_{e,k}$ and $\theta^{d}_{k}=\theta_{e}^{k}+\pi$.  As shown
at the end of Appendix C, the above conclusions also hold if the
approximation to $\widehat{G}^{k}_{0}(\vec{R})$ in
Eq.~(\ref{eq:GGGapprox}) is not made.

\begin{figure}[tbh]
\centering\includegraphics[height=15cm,width=11.7cm]{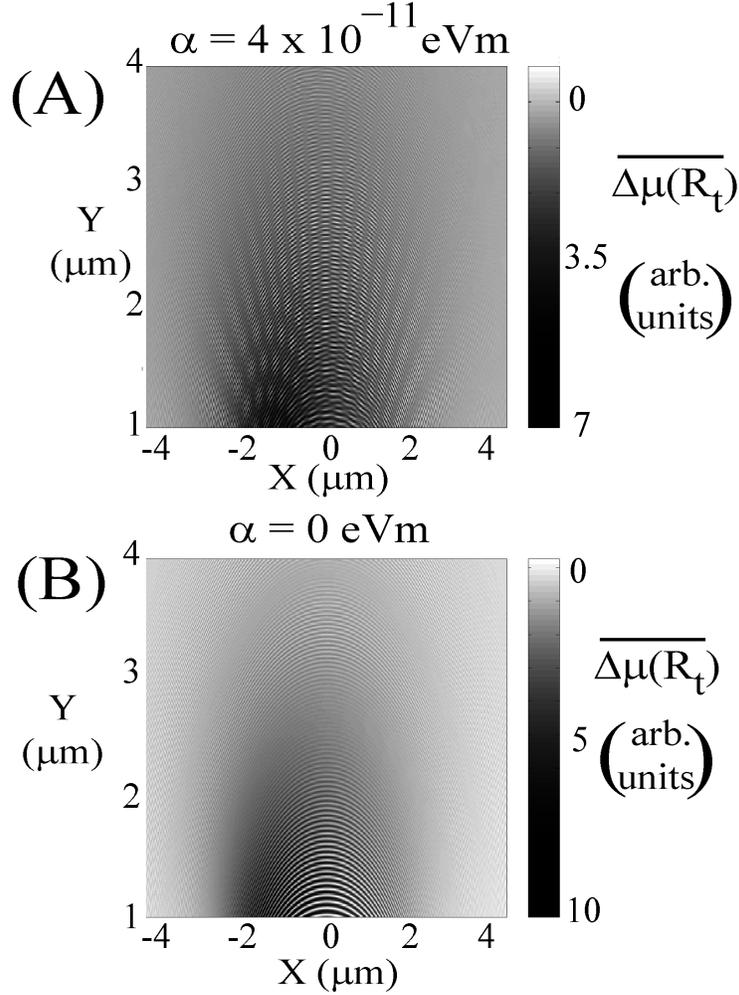}
\caption{Simulation of $\overline{\Delta\mu(\vec{R}_{t})}$
[Eq.~(\ref{eq:gensit})] for an unpolarized beam of electrons
injected from an emitter located at
$\vec{r}_{e}=1.5\,\mu\text{m}\widehat{X}$ and observed at a detector
located at $\vec{r}_{d}=-1.5\,\mu\text{m}\widehat{X}$ (A) with and
(B) without Rashba spin-orbit coupling for the same random
configuration of scatterers. A scatterer density of 20 scatterers
per $\mu\text{m}^{2}$ was chosen; each scatterer was modeled as a
cylindrical barrier/well of radius 3 nm with the height/depth of the
potential randomly chosen between $\pm$.04 eV. The tip was modeled
as a hard disc of radius 3 nm.  In (A), the amplitude of
$\overline{\Delta\mu(\vec{R}_{t})}$ is modulated by the Rashba
spin-orbit interaction, due to the fact that electrons traveling
from the emitter to the detector undergo net spin rotations.  In
(B), the expected elliptical fringes are observed. The same,
arbitrary scale for $\overline{\Delta\mu(\vec{R}_{t})}$ was used in
both (A) and (B).
 The following parameters were used in the simulation: $m^{*}=0.022
m_{0}$, $T=3$ K, $E_{F}=16$ meV, $\alpha=4\times10^{-11}$ eVm.}
\label{fig:figure3}\end{figure}

Consider the case explicitly illustrated in Figure \ref{fig:figure2}
where the detector and the emitter are two distinct entities. In
this case, an electron does not traverse the same path back to the
detector, and the effects of spin-orbit coupling do not average away
in the flux calculation. Performing the thermal averaging and spin
averaging [using Eqs. (\ref{eq:phi1})-(\ref{eq:phi2})],
$\overline{\Delta\mu(\vec{R}_{t})}$ can be written as:
\begin{eqnarray}
\overline{\Delta
\mu(\vec{R}_{t})}&=&\frac{2\epsilon}{\pi^{2}}\frac{\hbar\overline{k}_{F}}{m^{*}}\left(1+\frac{\pi^{2}}{12}\left(\frac{k_{B}T}{E_{F}}\right)^{2}\right)\overline{t}^{2}_{t}\frac{\cos(\theta^{d}_{t})}{r_{e,tip}r_{tip,d}}\nonumber\\
&+&\exp(\text{i}\overline{k}_{F}r_{S})\left(G_{1}(\vec{R}_{t})\exp(\text{i}k_{\alpha}r_{S})+G_{2}(\vec{R}_{t})\exp(-\text{i}k_{\alpha}r_{S})\right)\nonumber\\
&+&\exp(\text{i}\overline{k}_{F}r_{S})\left(G_{3}(\vec{R}_{t})\exp(\text{i}k_{\alpha}r_{D})+G_{4}(\vec{R}_{t})\exp(-\text{i}k_{\alpha}r_{D})\right)+h.c.
\label{eq:gensit}
\end{eqnarray}
where $r_{S}=r_{e,tip}+r_{tip,d}$ and $r_{D}=r_{tip,d}-r_{e,tip}$,
and where $G_{1}(\vec{R}_{t})$, $G_{2}(\vec{R}_{t})$,
$G_{3}(\vec{R}_{t})$, and $G_{4}(\vec{R}_{t})$ are nonoscillatory
functions of tip position which depend upon the particular
configuration of scatterers. From Eq.~(\ref{eq:gensit}), the
expected elliptical fringes spaced at $\overline{k}_{F}r_{S}=2n\pi$
with the detector and the emitter acting as the foci of the ellipse
are present;  however, the amplitude of these oscillations are now
modulated by the Rashba spin-orbit coupling.  Since the electron
trajectories from the emitter to the detector QPC are now
two-dimensional, the electron's spin will undergo a
trajectory-dependent spin rotation for each pathway between the
emitter and the detector QPC.  The interference between different
pathways will thus have an additional, spin-dependent modulation.
 Such an amplitude modulation is similar to the Elliot-Yafet model of
spin dephasing in the presence of spin-orbit
coupling\cite{Elliott54}. The first two oscillatory terms in
Eq.~(\ref{eq:gensit}), $G_{1}(\vec{R}_{t})$ and
$G_{2}(\vec{R}_{t})$, lead to elliptical amplitude modulations of
the regular fringes spaced at $k_{\alpha}r_{S}=2n\pi$, while the
$G_{3}$ and $G_{4}$ terms in Eq.~(\ref{eq:gensit}) lead to a
hyperbolic amplitude modulation spaced at $k_{\alpha}r_{D}=2n\pi$,
with the foci of the hyperbola again being the detector and the
emitter. Interference between the terms $G_{1}$ and $G_{3}$ and
between the terms $G_{2}$ and $G_{4}$ lead to fringes at
$k_{\alpha}r_{e,tip}=2n\pi$ for $r_{tip,d}=$constant. Likewise,
interference between the terms $G_{1}$ and $G_{4}$ and between the
terms $G_{2}$ and $G_{3}$ lead to fringes at
$k_{\alpha}r_{tip,d}=2n\pi$ for $r_{e,tip}=$constant. The presence
of all four types of fringe patterns leads to a checkered pattern in
$\overline{\Delta\mu(\vec{R}_{t})}$. This can be seen in the
calculation of $\overline{\Delta\mu(\vec{R}_{t})}$ in
Eq.~(\ref{eq:gensit}) which is shown in Fig. \ref{fig:figure3} (A).
In this simulation, the emitter was placed at
$\vec{r}_{e}=1.5\mu\,\text{m}\widehat{X}$ and the detector was
placed $3\mu$m away at $\vec{r}_{d}=-1.5\mu\,\text{m}\widehat{X}$.
In addition, the thermal average of $\Delta\mu(\vec{R}_{t})$ was
evaluated by numerically integrating Eq.~(\ref{eq:thermie}) over the
interval $E_{F}\pm 6k_{B}T$, and the following parameters were used
(similar to the parameters found for InAs\cite{Grundler00}):
$m^{*}=0.022 m_{0}$ (where $m_{0}$ is the free electron mass),
$E_{F}=$ 16 meV, $T=3$K, and
$\alpha=4\times10^{-11}\text{eV}\text{m}$ which gives a spin
rotation length (i.e., the length required to rotate the spin by
$180^{\circ}$ of $l_{\pi}=\pi\hbar^{2}/(2m^{*}\alpha)$) of 134 nm
(For comparison, spin rotation lengths of $l_{\pi}\approx
1.8\mu\text{m}$ were found for heterostructures of GaAs/AlGaAs in a
past study\cite{Miller03}).  Scatterers were randomly placed in the
region $[Y,X]=$(0 $\mu\text{m}$, 6 $\mu$m)$\times(-6\,\mu$m, 6
$\mu$m) with a scatterer density of 20 scatterers per
$\mu\text{m}^{2}$.  All scatterers were modeled as cylindrical wells
or barriers, with the well depths randomly chosen between $\pm .04$
eV.  The coupling constants, $\overline{t}_{k}$, were evaluated
using Eq.~(\ref{eq:params}) and the results in Appendix A. The
radius of the barriers/wells were all taken to be $a=3$ nm, which
gave $\overline{k}_{F}a\approx .3$ so that a model of ``s"-wave
scatterers could be used.  The tip was modeled as a hard disc (i.e.,
infinite barrier) with the radius of the tip chosen to be $a=3$ nm.
Although the width of the actual depletion area induced by the tip
in the 2DEG is probably on the order of 100 nm, the above radius was
chosen to be consistent with the ``s"-wave model used in the
calculations (further studies incorporating higher partial wave
scattering in the presence of Rashba coupling are currently being
carried out and will be addressed at a later time).  Figure
\ref{fig:figure3}(A) is a typical result from the calculations
performed on numerous scatterer configurations. Besides the
kinematical elliptical fringes spaced at
$\overline{k}_{F}r_{S}=2n\pi$, the hyperbolic and elliptical
modulations are clearly present in Figure \ref{fig:figure3} (A),
along with the circular-like fringes about the emitter and the
detector, which leads to a checkered pattern. \clearpage
\clearpage

\begin{figure}[tbp]
\centering \includegraphics[width=10.7cm]{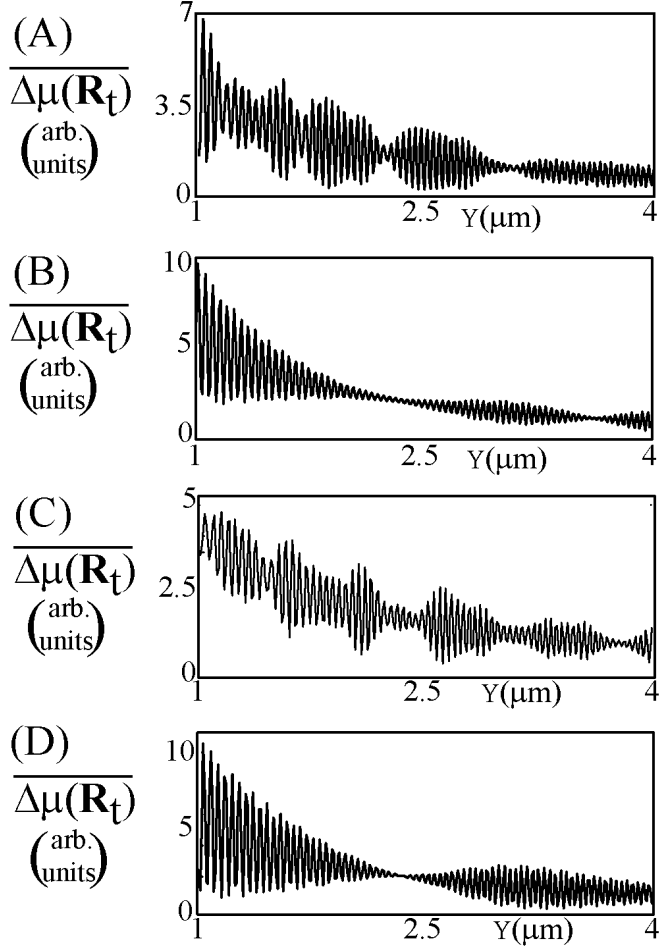} \caption{Slices
of $\overline{\Delta\mu(\vec{R}_{t})}$ through the detector at
$\vec{r}_{d}=-1.5\mu$m $\widehat{X}$. The plots shown in (A) and (B)
come from Figure \ref{fig:figure3}. Modulations are clearly evident
in (A) due to spin-orbit coupling. Note also that the intensity of
the fringes is larger in the absence of spin-orbit coupling [(B) vs.
(A)]. However, the fringe intensities also depend upon the
configuration of the random impurities.  (C) and (D) show the same
slice of $\overline{\Delta\mu(\vec{R}_{t})}$ for a different
configuration of scatterers.  Note again that modulations due to
Rashba coupling are present in (C) and not in (D).}
\label{fig:figure4}
\end{figure}

 For a comparison, simulations were
also performed for the same scattering configurations and coupling
constants but without the Rashba spin-orbit coupling ($\alpha=0$
eVm). The fermi energy of these simulations, $E_{F}^{'}$, was chosen
to be slightly higher in energy than $E_{F}$ in Figure
\ref{fig:figure3}(A) so that the magnitude of the fermi vectors was
the same for both simulations, i.e.,
$\overline{k}_{F}^{'}=\overline{k}_{F}$. As expected, only regular
elliptical fringes about the emitter and detector are shown in
Figure \ref{fig:figure3}(B), with any resulting modulation arising
from the particular scatterer configuration. Note that the
intensities of the fringes are larger when the tip is near to the
detector than the corresponding fringes in the presence of Rashba
coupling. Figure \ref{fig:figure4}(A) and (B) demonstrates this more
clearly by plotting a slice of $\overline{\Delta\mu(\vec{R}_{t})}$
[shown in Figure \ref{fig:figure3}(A)and (B)] along the
$\widehat{Y}$-axis, passing through the detector.  Near the
detector, the magnitude of $\overline{\Delta\mu(\vec{R}_{t})}$ is
greater in the absence of Rashba coupling.  However, the magnitudes
of $\overline{\Delta\mu(\vec{R}_{t})}$ with and without spin-orbit
coupling are comparable far away from the detector. This is due to
the fact that far away from the detector and emitter, the scattered
wave functions can no longer resolve the detector and the emitter,
i.e.,
$\widehat{U}(k_{\alpha}r_{k,d},\theta_{d}^{k})\widehat{U}(k_{\alpha}r_{e,k},\theta_{e}^{k})\approx\widehat{1}$,
which makes $\overline{\Delta\mu(\vec{R}_{t})}$ independent of
$\alpha$.  It must be stressed, however, that although the form of
the fringe pattern is robust to scatterer configurations, the
overall intensity does depend on the scattering configuration.
 Figure \ref{fig:figure4}(C) and \ref{fig:figure4}(D) give the same slice through a
system with a different set of scatterers.
\subsection{Two Scatterer solution}
\begin{figure}[tbp]
\centering \includegraphics[width=10.7cm]{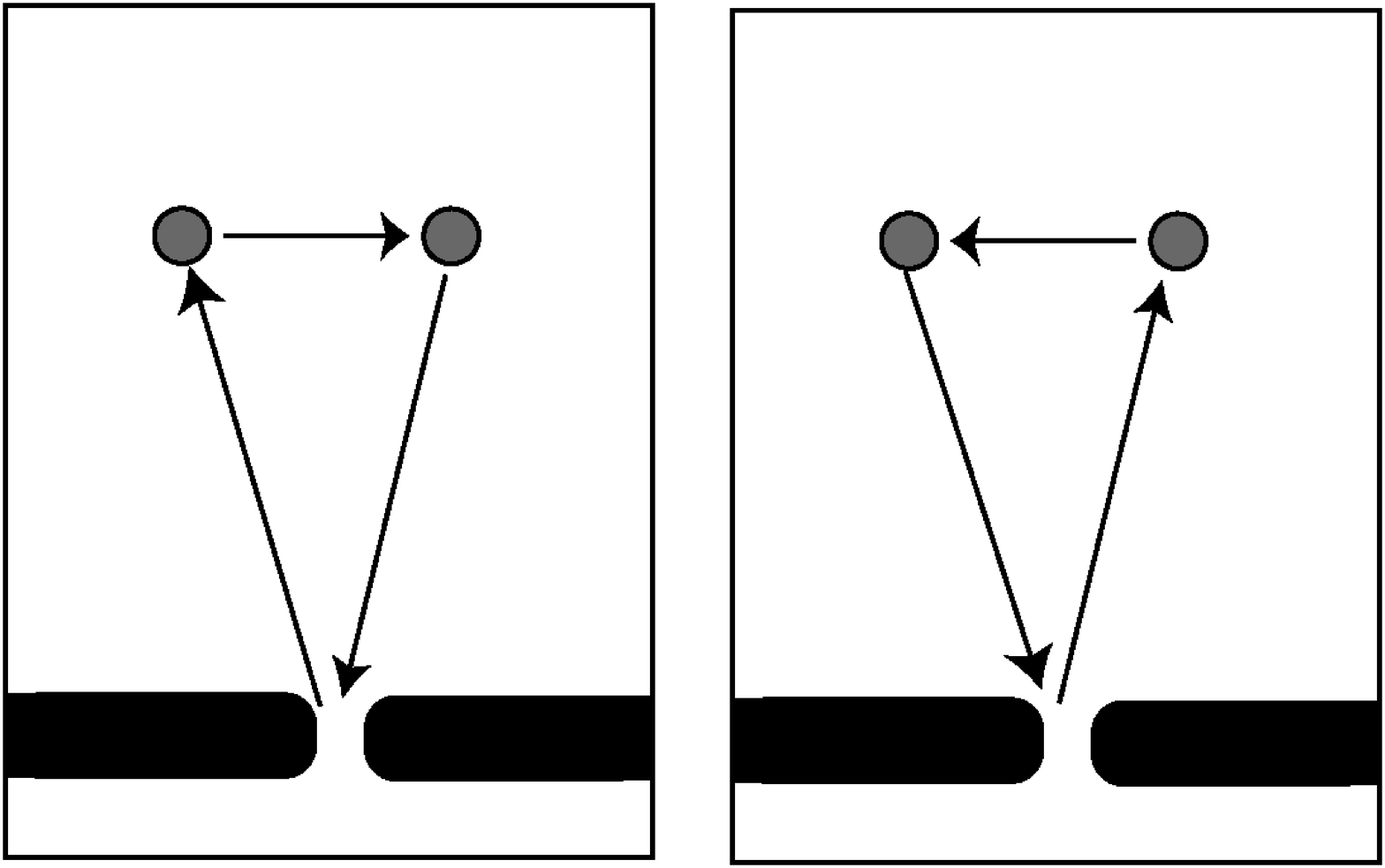}
\caption{Possible multiple scattering trajectories which result in a
net spin rotation applied to spins which make round trips from the
emitter back to the emitter.  Since the path is two-dimensional, a
net spin rotation results, even when the detector and the emitter
are the same.} \label{fig:figure5}
\end{figure}

\begin{figure}[tbh]
\centering
 \includegraphics[width=10cm]{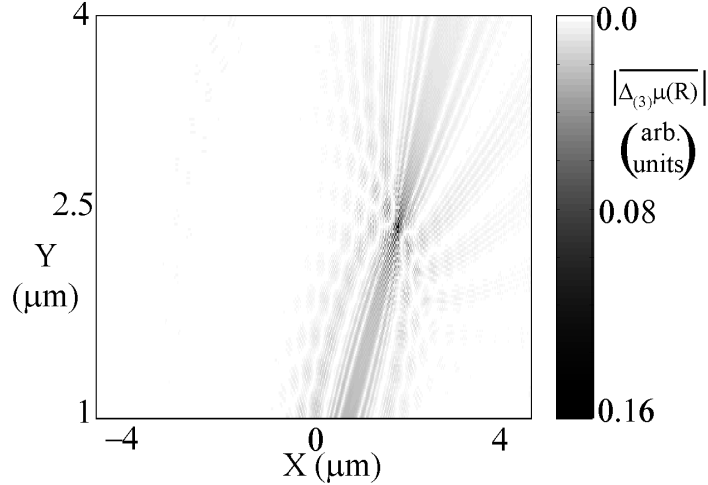}
\caption{Calculation of $|\overline{\Delta_{(3)}\mu(\vec{R}_{t})}|$
in Eq.~(\ref{eq:dmu3}) for a fixed hard disc scatterer of radius 3
nm located at
$\vec{r}_{S}=1.7\,\mu\text{m}\widehat{X}+2.4\,\mu\text{m}\widehat{Y}$.
The modulations due to Rashba spin-orbit coupling are seen, similar
to those shown in Figure \ref{fig:figure3}.  These modulations
mainly result from the interference between the trajectories shown
in Figure \ref{fig:figure5} and the single scattering trajectories
from the random impurities. The flux scale and parameters used in
the calculation of $|\overline{\Delta_{(3)}\mu(\vec{R}_{t})}|$ are
the same as those given in Figure \ref{fig:figure3}.
 } \label{fig:figure6}
\end{figure}

As mentioned earlier, if there is only one QPC, the effects of
spin-orbit coupling are not observed if the electron trajectories
from and towards the QPC are purely one-dimensional. As shown in
Figure \ref{fig:figure5}, two-dimensional multiple-scattering
trajectories exist for electrons exiting and arriving at the
detector. Consider the case of two scatterers: a moveable scatterer
at $\vec{R}_{t}$ and a fixed scatterer at $\vec{r}_{s}$.  In
addition, both the tip and the fixed scatterer will be modeled as
being infinite potential barriers of radius $3$ nm.  Using
Eq.~(\ref{eq:finalcdown}), the value of the wave function at each of
the two scatterers is given by
\begin{eqnarray}
\label{eq:psirt}
\Psi(\vec{R}_{t})&=&\lambda_{s,t}\Phi(\vec{R}_{t})-\lambda_{s,t}G^{s}_{0}(\vec{R}_{t})\Phi(\vec{r}_{s})\\
\Psi(\vec{r}_{s})&=&\lambda_{s,t}\Phi(\vec{r}_{s})-\lambda_{s,t}G^{t}_{0}(\vec{r}_{s})\Phi(\vec{R}_{t})
\label{eq:psirs}
\end{eqnarray}
where $\lambda_{s,t}=(1-\text{Det}[G^{t}_{0}(\vec{r}_{s})])^{-1}$.
This includes all orders of scattering between the two scatterers
(i.e., any number of bounces between the two scatterers).  The
change in current as a function of $\vec{R}_{t}$ in the presence of
the fixed scatterer at $\vec{r}_{s}$ and a random configuration of
weak scatterers can be found by inserting Eq.~(\ref{eq:psirt}) and
Eq.~(\ref{eq:psirs}) into Eq.~(\ref{eq:gencond}) and performing the
thermal average using Eq.~(\ref{eq:thermie}). Since
$\overline{t}<<1$ in the ``s"-wave limit, it is useful to expand
$\overline{\Delta \mu(\vec{R}_{t})}$ in powers of $\overline{t}$.
The single scattering contribution has already been discussed, and
is given in Eq.~(\ref{eq:fluxsame}), which is order
$\overline{t}^{2}$. The next term of order $\overline{t}^{3}$, which
involves the interference between the trajectories shown in Fig.
\ref{fig:figure5} and the single scattering trajectories, can be
written as:
\begin{eqnarray}
\overline{\Delta_{(3)}\mu(\vec{R}_{t})}&=&\exp(\text{i}\overline{k}\widetilde{R}_{S})\left(K_{1}(\vec{R}_{t})\exp(\text{i}k_{\alpha}\widetilde{R}_{S})+K_{2}(\vec{R}_{t})\exp(\text{i}k_{\alpha}\widetilde{R}_{D})\right)\nonumber\\
&+&\exp(\text{i}\overline{k}\widetilde{R}_{S})\left(K_{3}(\vec{R}_{t})\exp(-\text{i}k_{\alpha}\widetilde{R}_{S})+K_{4}(\vec{R}_{t})\exp(-\text{i}k_{\alpha}\widetilde{R}_{D})\right)\nonumber\\
&+&\exp(\text{i}\overline{k}\widetilde{R}_{D})\left(L_{1}(\vec{R}_{t})\exp(\text{i}k_{\alpha}\widetilde{R}_{S})+L_{2}(\vec{R}_{t})\exp(\text{i}k_{\alpha}\widetilde{R}_{D})\right)\nonumber\\
&+&\exp(\text{i}\overline{k}\widetilde{R}_{D})\left(L_{3}(\vec{R}_{t})\exp(-\text{i}k_{\alpha}\widetilde{R}_{S})+L_{4}(\vec{R}_{t})\exp(-\text{i}k_{\alpha}\widetilde{R}_{D})\right)+h.c\nonumber\\
\label{eq:dmu3}
\end{eqnarray}
where $\widetilde{R}_{S}=r_{tip,d}+r_{s,tip}$ and
$\widetilde{R}_{D}=r_{tip,d}-r_{s,tip}$.  The functions
$K_{1}(\vec{R}_{t}),K_{2}(\vec{R}_{t}),K_{3}(\vec{R}_{t}),$ and
$K_{4}(\vec{R}_{t})$ (which mostly represent the interference
between the impurity single scattering events and the trajectories
shown in Fig. \ref{fig:figure5}) are nonoscillatory functions of
$\vec{R}_{t}$ which depend upon the configuration of random
scatterers, whereas the functions $L_{1}(\vec{R}_{t}),
L_{2}(\vec{R}_{t}), L_{3}(\vec{R}_{t})$ and $L_{4}(\vec{R}_{t})$
(which represent the interference between the single scattering
trajectories (for the tip and the fixed impurity) and the
multiple-scattering trajectories shown in Fig. \ref{fig:figure5})
are nonoscillatory functions of $\vec{R}_{t}$ which only depend upon
the position of the fixed scatterer at $\vec{r}_{s}$. Figure
\ref{fig:figure6} shows a simulation of
$|\overline{\Delta_{(3)}\mu(\vec{R}_{t})}|$ in Eq.~(\ref{eq:dmu3})
for a fixed, hard disc scatterer of radius 3 nm located at
$\vec{r}_{S}=1.7\,\mu\text{m}\widehat{X}+2.4\,\mu\text{m}\widehat{Y}$.
The detector/emitter QPC was placed at
$(X,Y)=(0\,\mu\text{m},\,0\,\mu\text{m})$.  Fig. \ref{fig:figure6}
represents the evaluation of Eq.~(\ref{eq:dmu3}) for the following
parameters:  $T=3$ K, $m^{*}=0.022 m_{0}$, $\alpha=4\times10^{-11}$
eVm, and $E_{F}=16$ meV (the same parameters as those given in
Figure \ref{fig:figure3}).  The random impurities were again modeled
as cylindrical wells/barriers of radius $3$ nm with depth/height
randomly chosen between $\pm .04$ eV and with a scatterer density of
$20$ scatterers per $\mu\text{m}^{2}$. Modulations due to spin-orbit
coupling are again present, as predicted in Eq.~(\ref{eq:dmu3}).

\begin{figure}[tbp]
\centering \includegraphics[width=10.7cm]{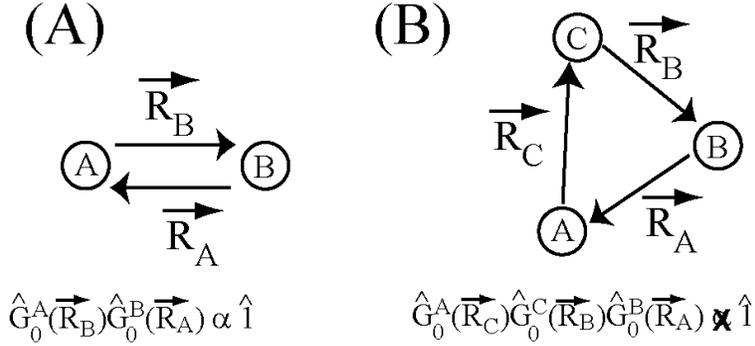}
\caption{Possible higher-order scattering processes.  For two
scatterers shown in (A), if the electron bounces between scatterer A
and B, no net spin rotation results.  For three or more scatterers,
however, net spin rotation can occur.  Such a possibility is shown
in (B), where an electron traveling from A to C to B to A undergoes
a spin rotation.} \label{fig:figure7}
\end{figure}

 If the scattering amplitudes, $\overline{t}$,
become large, then higher-orders (i.e., multiple bounces) must also
be included.  The interference between these different trajectories
can lead to resonances induced by the scattering configuration.  For
the trajectories shown in Figure \ref{fig:figure5}, however, no
resonances due to spin rotation can be generated, since no net spin
rotation is generated if the particle bounces from scatterer $A$ to
scatterer $B$ and back to scatterer $A$ again, as shown in Figure
\ref{fig:figure7}(A).  The lack of spin rotation for such
trajectories can be seen using the exact form of
$\widehat{G}^{k}_{0}(\vec{R})$ given in Eq.~(\ref{eq:GGG}) as
follows:
\begin{eqnarray}
\widehat{G}^{A}_{0}(\vec{R}_{B})\widehat{G}^{B}_{0}(\vec{R}_{A})\propto\widehat{1}
\label{eq:refocus}
\end{eqnarray}
However, for three or more scatterers(as shown in Figure
\ref{fig:figure7}(B)), there exist trajectories which will give a
net spin rotation.  Calculating possible interference effects
between multiple scattering trajectories requires using higher
partial waves [Eq.~(\ref{eq:Toperator})] than the simple ``s"-wave
scattering models studied mostly in this paper, and will be
investigated in the future.
\section{Conclusions}
In this paper, a partial wave expansion for scattering from a
cylindrical potential in the presence of Rashba spin-orbit coupling
was developed and was used to construct an operator
$\widehat{T}_{k}$ which generates the scattered wave from the
incident wave at scatterer $k$. This allowed for the development of
point scattering models beyond the ``s"-wave limit.  The often
studied ``s"-wave scattering from delta-function potentials are
shown to be different from the ``s"-wave models developed in this
work due to the fact that cylindrical wells/barriers do not scatter
in the delta-function limit;  however, both models give the same
qualitative results for the flux calculations presented in this
work. Although only ``s"-waves were discussed herein, extensions to
higher partial wave scattering can be readily performed using the
formalism developed in this work. Additionally, the operator
$\widehat{T}_{k}$ and the calculated Green's function,
$\widehat{G}_{\pm}(\vec{r}_{1},\vec{r}_{2},E)$, can be used to apply
all the scattering theory machinery to study 2DEG confined in a
variety of geometries.

The Rashba spin-orbit coupling was shown to generate additional
interference fringes in possible electron imaging experiments which
were produced using a moveable scanning probe microscope tip. In the
single-scattering limit, spin-orbit coupling doesn't produce any
modulation in the observed flux using a single quantum point
contact.  This is due to the fact that no net spin rotation is
generated from effective one-dimensional trajectories which start
and end at the same location.  If the injected current through a
separate quantum point contact is measured instead, interference
effects due to Rashba coupling are observed from the various
two-dimensional trajectories from the emitter to the detector.  This
is due to the noncommutation of the resulting spin rotations along
the trajectory, which results in spin-orbit-related interference
effects. These interference effects are similar to the Elliott-Yafet
mechanism of spin dephasing observed in electron systems. If
multiple-scattering effects are also included, a single quantum
point contact can again be used to observe the spin interference
caused by the Rashba spin-orbit coupling.

In the future, calculations involving higher partial waves and
stronger scattering will be performed in order to look for possible
spin resonances resulting from interference between the various
multiple scattering trajectories.  In addition, more realistic
simulations of the scattering induced by a scanning probe microscope
tip, requiring other partial waves in addition to the ``s"-waves,
will be performed. Finally, the multiple-scattering theory presented
in this work can also be used to study scattering and polarization
profiles generated in quantum wires where phase coherence effects
between the scattered waves can now be fully taken into account.
\begin{acknowledgments}We would like to thank the referees for their
comments, in particular, referee 2 for pointing out reference 27 in
regards to evaluating Eq. (\ref{eq:thermie}). This work was
supported at Harvard University by the Nanoscale Science and
Engineering Center (NSF Grant No. PHY-0117795), and by NSF Grant No.
CHE-0073544.
\end{acknowledgments}
\appendix
\section{Solution for scattering from a barrier/well}
For the problem of the square well/barrier centered about
$\vec{r}_{k}$, Eq.~(\ref{eq:Cont1}) and Eq.~(\ref{eq:Cont2}) require
the various $f^{\pm}_{l}$ and $d^{\pm}_{l}$ for
$|\vec{r}-\vec{r}_{k}|=a$ to satisfy the following equations:
\begin{eqnarray}
 \lambda^{\pm}_{l}J_{l}(k_{\pm}a)+\sqrt{k_{1}}f^{\pm
1}_{l}H_{l}(k_{1}a)+\sqrt{k_{2}}f^{\pm
2}_{l}H_{l}(k_{2}a)&=&\sqrt{\kappa_{1}}d^{\pm
1}_{l}J_{l}(\kappa_{1}a)+\sqrt{\kappa_{2}}d^{\pm 2}_{l}J_{l}(\kappa_{2}a)\nonumber\\
\pm \lambda^{\pm}_{l}J_{l-1}(k_{\pm}a)+\sqrt{k_{1}}f^{\pm
1}_{l}H_{l-1}(k_{1}a)-\sqrt{k_{2}}f^{\pm
2}_{l}H_{l-1}(k_{2}a)&=&\sqrt{\kappa_{1}}d^{\pm
1}_{l}J_{l-1}(\kappa_{1}a)-\sqrt{\kappa_{2}}d^{\pm 2}_{l}J_{l-1}(\kappa_{2}a)\nonumber\\
\lambda^{\pm}_{l}k_{\pm}J^{'}_{l}(k_{\pm}a)+k^{3/2}_{1}f^{\pm
1}_{l}H^{'}_{l}(k_{1}a)+k^{3/2}_{2}f^{\pm
2}_{l}H^{'}_{l}(k_{2}a)&=&\kappa^{3/2}_{1}d^{\pm
1}_{l}J^{'}_{l}(\kappa_{1}a)+\kappa^{3/2}_{2}d^{\pm 2}_{l}J^{'}_{l}(\kappa_{2}a)\nonumber\\
\pm k_{\pm}\lambda^{\pm}_{l}J^{'}_{l-1}(k_{\pm}a)+k^{3/2}_{1}f^{\pm
1}_{l}H^{'}_{l-1}(k_{1}a)-k^{3/2}_{2}f^{\pm
2}_{l}H^{'}_{l-1}(k_{2}a)&=&\kappa^{3/2}_{1}d^{\pm
1}_{l}J^{'}_{l-1}(\kappa_{1}a)-\kappa^{3/2}_{2}d^{\pm 2}_{l}J^{'}_{l-1}(\kappa_{2}a)\nonumber\\
\label{eq:matchingcond}
\end{eqnarray}
where\begin{eqnarray}
\lambda^{\pm}_{l}&=&2\text{i}^{l}\exp(\text{i}\vec{k}_{\pm}\cdot\vec{r}_{k})\exp(-\text{i}l\theta_{0})\end{eqnarray}
Using Eq.~(\ref{eq:matchingcond}), the various values for the
coefficients, $d^{\pm 1}_{l},d^{\pm 2}_{l},f^{\pm 1}_{l}$, and
$f^{\pm 2}_{l}$ can be found.  In order to simplify the presentation
of the solution to Eq.~(\ref{eq:matchingcond}), the following
functions will be introduced to simplify the solutions:
\begin{eqnarray}
\Delta_{l}(p,q,a)&=&J_{l}(pa)J_{l-1}(qa)+J_{l}(qa)J_{l-1}(pa)\nonumber\\
\Delta\Delta_{l}(p,q,a)&=&J'_{l}(pa)J'_{l-1}(qa)+J'_{l}(qa)J'_{l-1}(pa)\nonumber\\
g^{b}_{l}(p,q,a)&=&J_{l-1}(qa)H_{l}(pa)+(-1)^{b}J_{l}(qa)H_{l-1}(pa)\nonumber\\
G^{b}_{l}(p,q,r,a)&=&\frac{p\Delta_{l}(q,r,a)}{q\Delta\Delta_{l}(q,r,a)}\left(J'_{l-1}(ra)H'_{l}(pa)+(-1)^{b}
J^{'}_{l}(ra)H^{'}_{l-1}(pa)\right)-g^{b}(p,r,a)\nonumber\\
F^{b}_{l}(k_{\pm},q,r,a)&=&\left(J_{l-1}(ra)J_{l}(k_{\pm}a)\pm(-1)^{b}
J_{l-1}(k_{\pm}a)J_{l}(ra)\right)\nonumber\\
&-&\frac{k_{\pm}\Delta_{l}(q,r,a)}{q\Delta\Delta_{l}(q,r,a)}\left(J'_{l-1}(ra)J'_{l}(k_{\pm}a)\pm(-1)^{b}
J^{'}_{l-1}(k_{\pm}a)J^{'}_{l}(ra)\right)
\end{eqnarray}
The solution to the above equations can be written as
\begin{eqnarray}
f^{\pm
1}_{l}&=&\lambda^{\pm}_{l}\frac{F^{2}_{l}(k_{\pm},\kappa_{1},\kappa_{2},a)G^{2}_{l}(k_{2},\kappa_{2},\kappa_{1},a)-F^{1}_{l}(k_{\pm},\kappa_{2},\kappa_{1},a)G^{1}_{l}(k_{2},\kappa_{1},\kappa_{2},a)}
{\sqrt{k_{1}}\left(G^{2}_{l}(k_{1},\kappa_{1},\kappa_{2},a)G^{2}_{l}(k_{2},\kappa_{2},\kappa_{1},a)-G^{1}_{l}(k_{1},\kappa_{2},\kappa_{1},a)G^{1}_{l}(k_{2},\kappa_{1},\kappa_{2},a)\right)}\nonumber\\
&=&\lambda^{\pm}_{l}\frac{\widetilde{f}^{\pm 1}_{l}}{\sqrt{k_{1}}}\nonumber\\
 f^{\pm
2}_{l}&=&\lambda^{\pm}_{l}\frac{F^{2}_{l}(k_{\pm},\kappa_{1},\kappa_{2},a)G^{1}_{l}(k_{1},\kappa_{2},\kappa_{1},a)-F^{1}_{l}(k_{\pm},\kappa_{2},\kappa_{1},a)G^{2}_{l}(k_{1},\kappa_{1},\kappa_{2},a)}
{\sqrt{k_{2}}\left(G^{1}_{l}(k_{2},\kappa_{1},\kappa_{2},a)G^{1}_{l}(k_{1},\kappa_{2},\kappa_{1},a)-G^{2}_{l}(k_{2},\kappa_{2},\kappa_{1},a)G^{2}_{l}(k_{1},\kappa_{1},\kappa_{2},a)\right)}\nonumber\\
&=&\lambda^{\pm}_{l}\frac{\widetilde{f}^{\pm 2}_{l}}{\sqrt{k_{2}}}\nonumber\\
 d^{\pm 1}_{l}&=&\lambda^{\pm}_{l}\frac{J_{l-1}(\kappa_{2}a)J_{l}(k_{\pm}a)\pm
J_{l}(\kappa_{2}a)J_{l-1}(k_{\pm}a)}{\sqrt{\kappa_{1}}\Delta(\kappa_{1},\kappa_{2},a)}\nonumber\\
&+&\frac{\lambda^{\pm}_{l}}{\Delta(\kappa_{1},\kappa_{2},a)}\frac{F^{2}_{l}(k_{\pm},\kappa_{1},\kappa_{2},a)(g^{2}_{l}(k_{1},\kappa_{2},a)G^{2}_{l}(k_{2},\kappa_{2},\kappa_{1},a)-g^{1}(k_{2},\kappa_{2})G^{1}_{l}(k_{1},\kappa_{2},\kappa_{1},a))}
{\sqrt{\kappa_{1}}\left(G^{2}_{l}(k_{1},\kappa_{1},\kappa_{2},a)G^{2}_{l}(k_{2},\kappa_{2},\kappa_{1},a)-G^{1}_{l}(k_{1},\kappa_{2},\kappa_{1},a)G^{1}_{l}(k_{2},\kappa_{1},\kappa_{2},a)\right)}\nonumber\\
&-&\frac{\lambda^{\pm}_{l}}{\Delta(\kappa_{1},\kappa_{2},a)}\frac{F^{1}_{l}(k_{\pm},\kappa_{2},\kappa_{1},a)(g^{2}_{l}(k_{1},\kappa_{2},a)G^{1}_{l}(k_{2},\kappa_{1},\kappa_{2},a)-g^{1}(k_{2},\kappa_{2})G^{2}_{l}(k_{1},\kappa_{1},\kappa_{2},a))}
{\sqrt{\kappa_{1}}\left(G^{2}_{l}(k_{1},\kappa_{1},\kappa_{2},a)G^{2}_{l}(k_{2},\kappa_{2},\kappa_{1},a)-G^{1}_{l}(k_{1},\kappa_{2},\kappa_{1},a)G^{1}_{l}(k_{2},\kappa_{1},\kappa_{2},a)\right)}\nonumber\\
d^{\pm
2}_{l}&=&\lambda^{\pm}_{l}\frac{J_{l-1}(\kappa_{1}a)J_{l}(k_{\pm}a)\mp
J_{l}(\kappa_{1}a)J_{l-1}(k_{\pm}a)}{\sqrt{\kappa_{2}}\Delta(\kappa_{1},\kappa_{2},a)}\nonumber\\
&+&\frac{\lambda^{\pm}_{l}}{\Delta(\kappa_{1},\kappa_{2},a)}\frac{F^{2}_{l}(k_{\pm},\kappa_{1},\kappa_{2},a)(g^{1}_{l}(k_{1},\kappa_{1},a)G^{2}_{l}(k_{2},\kappa_{2},\kappa_{1},a)+g^{2}(k_{2},\kappa_{1})G^{1}_{l}(k_{1},\kappa_{2},\kappa_{1},a))}
{\sqrt{\kappa_{2}}\left(G^{2}_{l}(k_{1},\kappa_{1},\kappa_{2},a)G^{2}_{l}(k_{2},\kappa_{2},\kappa_{1},a)-G^{1}_{l}(k_{1},\kappa_{2},\kappa_{1},a)G^{1}_{l}(k_{2},\kappa_{1},\kappa_{2},a)\right)}\nonumber\\
&-&\frac{\lambda^{\pm}_{l}}{\Delta(\kappa_{1},\kappa_{2},a)}\frac{F^{1}_{l}(k_{\pm},\kappa_{2},\kappa_{1},a)(g^{1}_{l}(k_{1},\kappa_{1},a)G^{1}_{l}(k_{2},\kappa_{1},\kappa_{2},a)+g^{2}(k_{2},\kappa_{1})G^{2}_{l}(k_{1},\kappa_{1},\kappa_{2},a))}
{\sqrt{\kappa_{2}}\left(G^{2}_{l}(k_{1},\kappa_{1},\kappa_{2},a)G^{2}_{l}(k_{2},\kappa_{2},\kappa_{1},a)-G^{1}_{l}(k_{1},\kappa_{2},\kappa_{1},a)G^{1}_{l}(k_{2},\kappa_{1},\kappa_{2},a)\right)}\nonumber\\
\end{eqnarray}
It is useful to consider the limiting case of a hard disc, i.e.,
$V_{0}\rightarrow\infty$.  In this case,
\begin{eqnarray}
\kappa_{1}&=&\frac{m^{*}\alpha}{\hbar^{2}}+\text{i}\sqrt{\frac{2m^{*}|V_{0}|}{\hbar^{2}}}\nonumber\\
\kappa_{2}&=&-\frac{m^{*}\alpha}{\hbar^{2}}+\text{i}\sqrt{\frac{2m^{*}|V_{0}|}{\hbar^{2}}}
\end{eqnarray}
with $|V_{0}|\rightarrow \infty$.  In this limit, the coefficients
are given by:\begin{eqnarray} f^{\pm
1}_{l}&=&-\frac{\lambda^{\pm}_{l}}{\sqrt{k_{1}}}\frac{J_{l}(k_{\pm}a)H_{l-1}(k_{2}a)\pm
J_{l-1}(k_{\pm}a)H_{l}(k_{2}a)}{H_{l}(k_{1}a)H_{l-1}(k_{2}a)+H_{l}(k_{2}a)H_{l-1}(k_{1}a)}\nonumber\\
f^{\pm
2}_{l}&=&-\frac{\lambda^{\pm}_{l}}{\sqrt{k_{2}}}\frac{J_{l}(k_{\pm}a)H_{l-1}(k_{1}a)\mp
J_{l-1}(k_{\pm}a)H_{l}(k_{1}a)}{H_{l}(k_{1}a)H_{l-1}(k_{2}a)+H_{l}(k_{2}a)H_{l-1}(k_{1}a)}
\label{eq:asymptotcoupl}
\end{eqnarray}
Note also that in the opposite limit of an infinite well,
$V_{0}\rightarrow -\infty$, the expressions for $f^{\pm 1}_{l}$ and
$f^{\pm 2}_{l}$ are the same as those given in
Eq.~(\ref{eq:asymptotcoupl}).
\section{Construction of $\widehat{D}_{l}$}

For an arbitrary plane wave state specifed by energy $E$ and with
incident momentum vectors making an angle of $\theta_{0}$ with
respect to the $\widehat{Y}$-axis,\begin{eqnarray}
\binom{\Psi^{\uparrow}(\vec{R})}{\Psi_{\downarrow}(\vec{R})}&=&\left(\frac{A}{\sqrt{2}}\exp(\text{i}\vec{k}_{1}\cdot\vec{R})\binom{1}{\exp(-\text{i}\theta_{0})}+\frac{B}{\sqrt{2}}\exp(\text{i}\vec{k}_{2}\cdot\vec{R})\binom{1}{-\exp(-\text{i}\theta_{0})}\right)
\label{eq:Psibib}
\end{eqnarray}
an operator $\widehat{D}_{l}$ can be constructed such that
\begin{eqnarray}
\widehat{D}_{l}\binom{\Psi^{\uparrow}(\vec{R})}{\Psi^{\downarrow}(\vec{R})}&=&\exp(\text{i}l\theta_{0})\binom{\Psi^{\uparrow}(\vec{R})}{\Psi^{\downarrow}(\vec{R})}
\end{eqnarray}

First define the operator $\widehat{P}_{l}$ by
\begin{eqnarray}
\widehat{P}_{l}&=&\left(\frac{l}{|l|}\frac{\partial}{\partial
R_{X}}-\text{i}\frac{\partial}{\partial R_{Y}}\right)^{|l|}
\end{eqnarray}
with $\widehat{P}_{0}=1$.   Exponential functions of the form
$\exp(\text{i}\vec{k}\cdot\vec{R})$, where
$\vec{k}=k(\cos(\theta_{0})\widehat{Y}+\sin(\theta_{0})\widehat{X})$,
are eigenfunctions of $\widehat{P}_{l}$, where
\begin{eqnarray}
\widehat{P}_{l}\exp(\text{i}\vec{k}\cdot\vec{R})&=&k^{|l|}\exp(\text{i}l\theta_{0})\exp(\text{i}\vec{k}\cdot\vec{R})
\end{eqnarray}

The operator $\widehat{D}_{l}$ can be decomposed in terms of the
operators $\widehat{P}_{l}$ as follows:
\begin{eqnarray}
\widehat{D}_{l}&=&\left(\begin{array}{cc}a_{l}\widehat{P}_{l}&b_{l}\widehat{P}_{l+1}\\
c_{l}\widehat{P}_{l-1}&d_{l}\widehat{P}_{l}\end{array}\right)
\end{eqnarray}
where the coefficients, $a_{l},b_{l},c_{l}$ and $d_{l}$ need to be
determined.  Operating $\widehat{D}_{l}$ on $\Psi(\vec{R})$ in
Eq.~(\ref{eq:Psibib}):\begin{eqnarray}
\widehat{D}_{l}\binom{\Psi^{\uparrow}(\vec{R})}{\Psi^{\downarrow}(\vec{R})}&=&\frac{A\exp(\text{i}l\theta_{0})\exp(\text{i}\vec{k}_{1}\cdot\vec{R})}{\sqrt{2}}\binom{a_{l}k^{|l|}_{1}+b_{l}k^{|l+1|}_{1}}{\exp(-\text{i}\theta_{0})(c_{l}k^{|l-1|}_{1}+d_{l}k^{|l|}_{1})}\nonumber\\
&+&\frac{B\exp(\text{i}l\theta_{0})\exp(\text{i}\vec{k}_{2}\cdot\vec{R})}{\sqrt{2}}\binom{a_{l}k^{|l|}_{2}-b_{l}k^{|l+1|}_{2}}{-\exp(-\text{i}\theta_{0})(d_{l}k_{2}^{|l|}-c_{l}k^{|l-1|}_{2})}\nonumber\\
&=&\exp(\text{i}l\theta_{0})\binom{\Psi^{\uparrow}(\vec{R})}{\Psi^{\downarrow}(\vec{R})}
\end{eqnarray}
The various coefficients must therefore satisfy
\begin{eqnarray}
a_{l}k^{|l|}_{1}+b_{l}k^{|1+l|}_{1}&=&1\nonumber\\
a_{l}k^{|l|}_{2}-b_{l}k^{|1+l|}_{2}&=&1\nonumber\\
c_{l}k^{|l-1|}_{1}+d_{l}k^{|l|}_{1}&=&1\nonumber\\
d_{l}k^{|l|}_{2}-c_{l}k^{|l-1|}_{2}&=&1
\end{eqnarray}
which gives \begin{eqnarray}
a_{l}&=&\frac{k_{2}^{|l+1|}+k_{1}^{|l+1|}}{k_{2}^{|l+1|}k_{1}^{|l|}+k_{1}^{|l+1|}k_{2}^{|l|}}\nonumber\\
b_{l}&=&\frac{k^{|l|}_{2}-k^{|l|}_{1}}{k^{|l+1|}_{2}k^{|l|}_{1}+k^{|l+1|}_{1}k^{|l|}_{2}}\nonumber\\
c_{l}&=&\frac{k^{|l|}_{2}-k^{|l|}_{1}}{k^{|l-1|}_{2}k^{|l|}_{1}+k^{|l-1|}_{1}k^{|l|}_{2}}\nonumber\\
d_{l}&=&\frac{k_{2}^{|l-1|}+k_{1}^{|l-1|}}{k_{2}^{|l-1|}k_{1}^{|l|}+k_{1}^{|l-1|}k_{2}^{|l|}}
\end{eqnarray}

For an arbitrary eigenstate of $\widehat{H}_{0}$ [Eq.~(\ref{eq:Ho})]
with energy $E$,
\begin{eqnarray}
\binom{\Psi^{\uparrow}(\vec{R})}{\Psi^{\downarrow}(\vec{R})}&=&\int^{2\pi}_{0}\text{d}\theta_{0}\left(\frac{\Psi^{1}(\theta_{0})}{\sqrt{2}}e^{\vec{R}\cdot\vec{k}_{1}(\theta_{0})}\binom{1}{\exp(-\text{i}\theta_{0})}+\frac{\Psi^{2}(\theta_{0})}{\sqrt{2}}e^{\vec{R}\cdot\vec{k}_{2}(\theta_{0})}\binom{1}{-\exp(-\text{i}\theta_{0})}\right)\nonumber\\
&\equiv&\int^{2\pi}_{0}\text{d}\theta_{0}\Psi(\theta_{0})
\end{eqnarray}
Operating $\widehat{D}_{l}$ upon $\Psi(\vec{R})$ gives
\begin{eqnarray}
\widehat{D}_{l}\binom{\Psi^{\uparrow}(\vec{R})}{\Psi^{\downarrow}(\vec{R})}&=&\int^{2\pi}_{0}\text{d}\theta_{0}\exp(\text{i}l\theta_{0})\Psi(\theta_{0})
\end{eqnarray}
\section{The Green's Function in the Presence of Spin-Orbit
Coupling} The Green's function in the presence of Rashba spin-orbit
interaction has a simple form in momentum space and can be written
as
\begin{eqnarray}
\widehat{G}_{\pm}(E)&=&-\lim_{\epsilon\rightarrow
0}\frac{\widehat{1}}{\widehat{H}-E\pm
\text{i}\epsilon}\nonumber\\
&=&-\lim_{\epsilon\rightarrow
0}\frac{m^{*}}{(2\pi\hbar)^{2}}\int\text{d}\vec{k}\frac{|\vec{k}\rangle\langle\vec{k}|}{|\vec{k}|^{2}-\frac{2m^{*}|\vec{k}|\alpha}{\hbar^{2}}-\frac{2m^{*}E}{\hbar^{2}}\pm\text{i}\epsilon}\left(\widehat{1}+\widehat{\sigma}_{X}\cos(\phi_{\vec{k}})-\widehat{\sigma}_{Y}\sin(\phi_{\vec{k}})\right)\nonumber\\
&-&\lim_{\epsilon\rightarrow
0}\frac{m^{*}}{(2\pi\hbar)^{2}}\int\text{d}\vec{k}\frac{|\vec{k}\rangle\langle\vec{k}|}{|\vec{k}|^{2}+\frac{2m^{*}|\vec{k}|\alpha}{\hbar^{2}}-\frac{2m^{*}E}{\hbar^{2}}\pm\text{i}\epsilon}\left(\widehat{1}-\widehat{\sigma}_{X}\cos(\phi_{\vec{k}})+\widehat{\sigma}_{Y}\sin(\phi_{\vec{k}})\right)\nonumber\\
\label{eq:Gregform}
\end{eqnarray}
where  $\phi_{\vec{k}}$ is the angle $\vec{k}$ makes with the
respect to the $\widehat{Y}$-axis. The form of the Green's function
in Eq.~(\ref{eq:Gregform}) has been used in numerous studies of spin
dynamics in 2DEGS.  The position space representation of the Green's
function,
$\widehat{G}_{\pm}(\vec{r}_{1},\vec{r}_{2},E)=\langle\vec{r}_{1}|\widehat{G}_{\pm}(E)|\vec{r}_{2}\rangle$,
however, has not been used to the best of the authors' knowledge. In
the presence of a scattering potential, $V(\vec{r})$, the total wave
function with energy $E$ is related to
$\widehat{G}_{\pm}(\vec{r}_{1},\vec{r}_{2},E)$ and $V(\vec{r})$ by
the Lipmann-Schwinger equation:
\begin{eqnarray}
\Psi(\vec{R})&=&\Phi(\vec{R})+\int\text{d}\vec{r}\widehat{G}_{+}(\vec{R},\vec{r},E)V(\vec{r})\Psi(\vec{r})
\end{eqnarray}
where $\Phi(\vec{R})$ would be the wave function in the absence of
the scattering potential $V(\vec{r})$.

Before calculating $\widehat{G}_{\pm}(\vec{r}_{1},\vec{r}_{2},E)$,
it is worth noting that the Rashba Hamiltonian, $\widehat{H}_{0}$
[Eq.~(\ref{eq:Ho})], is invariant to combined rotations in spin and
space about the $\widehat{z}$ axis:
\begin{eqnarray}
\widehat{H}_{0}&=&\widehat{M}(\theta)\widehat{H}_{0}\widehat{M}^{\dagger}(\theta)\label{eq:property1}\\
\widehat{M}(\theta)&=&\exp\left(-\text{i}\frac{\theta}{\hbar}
\widehat{L}_{Z}\right)\exp\left(-\text{i}\frac{\theta}{2}\sigma_{Z}\right)\nonumber\\
&=&\widehat{R}(\theta)\widehat{Z}(\theta)
\end{eqnarray}
Due to the above symmetry, the Green's function is also
invariant:\begin{eqnarray}
\widehat{G}_{\pm}(E)&=&\widehat{M}(\theta)\widehat{G}_{\pm}(E)\widehat{M}^{\dagger}(\theta)
\label{eq:GFinv}
\end{eqnarray}
From Eq.~(\ref{eq:GFinv}), it follows that\begin{eqnarray} \langle
\vec{r}_{1}|\widehat{G}_{\pm}(E)|\vec{r}_{2}\rangle&=&\langle
\vec{r}_{1}|\widehat{M}^{\dagger}(\theta)\widehat{M}(\theta)\widehat{G}_{\pm}(E)\widehat{M}^{\dagger}(\theta)\widehat{M}(\theta)|\vec{r}_{2}\rangle\nonumber\\
&=&\widehat{Z}^{\dagger}(\theta)\langle
\widehat{R}(\theta)\vec{r}_{1}|\widehat{G}_{\pm}(E)|\widehat{R}(\theta)\vec{r}_{2}\rangle\widehat{Z}(\theta)
\label{eq:niceprop}
\end{eqnarray}

$\widehat{G}_{\pm}(\vec{r}_{1},\vec{r}_{2},E)$ can be written as
\begin{eqnarray}
\widehat{G}_{\pm}(\vec{r}_{1},\vec{r}_{2},E)&=&\lim_{\epsilon\rightarrow
0}\frac{-m^{*}}{(2\pi\hbar)^{2}}\int\text{d}\vec{k}\left(\frac{\exp\left(\vec{k}\cdot\vec{r}\right)\widehat{1}}{|\vec{k}|^{2}+\frac{2m^{*}\alpha
|k|}{\hbar^{2}}-\frac{2m^{*}E}{\hbar^{2}}\pm
\text{i}\epsilon}+\frac{\exp\left(\vec{k}\cdot\vec{r}\right)\widehat{1}}{|\vec{k}|^{2}-\frac{2\alpha
m^{*}|k|}{\hbar^{2}}-\frac{2m^{*}E}{\hbar^{2}}\pm
\text{i}\epsilon}\right)\nonumber\\
&+&\exp\left(\vec{k}\cdot\vec{r}\right)\left(\frac{\widehat{\sigma}_{X}\cos(\phi_{\vec{k}})-\widehat{\sigma}_{Y}\sin(\phi_{\vec{k}})}{|\vec{k}|^{2}-\frac{2m^{*}\alpha
|k|}{\hbar^{2}}-\frac{2m^{*}E}{\hbar^{2}}\pm
\text{i}\epsilon}-\frac{\widehat{\sigma}_{X}\cos(\phi_{\vec{k}})-\widehat{\sigma}_{Y}\sin(\phi_{\vec{k}})}{|\vec{k}|^{2}+\frac{2\alpha
m^{*}|k|}{\hbar^{2}}-\frac{2m^{*}E}{\hbar^{2}}\pm\text{i}\epsilon}\right)
\label{eq:Grrr}
\end{eqnarray}
where $\vec{r}=\vec{r}_{1}-\vec{r}_{2}$.  In evaluating
Eq.~(\ref{eq:Grrr}), it is advantageous to take $\vec{r}$ to be
along say the $\widehat{Y}$-axis.  If
$\vec{r}=r[\cos(\theta)\widehat{Y}+\sin(\theta)\widehat{X}]$, then
$\widehat{R}(\theta)|\vec{r}\rangle=|r\widehat{Y}\rangle$.
Therefore, using Eq.~(\ref{eq:niceprop}), Eq.~(\ref{eq:Grrr}) can be
written as:\begin{eqnarray}
\widehat{G}_{\pm}(\vec{r}_{1},\vec{r}_{2},E)&=&
\lim_{\epsilon\rightarrow
0}\frac{-m^{*}}{(2\pi\hbar)^{2}}\int^{2\pi}_{0}\text{d}\phi\int^{\infty}_{0}\text{d}k
\left(\frac{k\exp(\text{i}kr\cos(\phi))\widehat{1}}{(k-k_{1}\pm
\text{i}\epsilon)(k+k_{2})}+\frac{k\exp(\text{i}kr\cos(\phi))\widehat{1}}{(k+k_{1})(k-k_{2}\pm\text{i}\epsilon)}\right)\nonumber\\
&-&k\left(\frac{Z^{\dagger}(\theta)(\widehat{\sigma}_{X}\cos(\phi)-\widehat{\sigma}_{Y}\sin(\phi))Z(\theta)}{(k-k_1)(k+k_2\pm\text{i}\epsilon)}-\frac{Z^{\dagger}(\theta)(\widehat{\sigma}_{X}\cos(\phi)-\widehat{\sigma}_{Y}\sin(\phi))Z(\theta)}{(k+k_1\pm\text{i}\epsilon)(k-k_2)}\right)\nonumber\\
\label{eq:Gr}\end{eqnarray} where
\begin{eqnarray}
k_{1}&=&\frac{m^{*}\alpha}{\hbar^{2}}+\sqrt{\left(\frac{m^{*}\alpha}{\hbar^{2}}\right)^{2}+\frac{2m^{*}E}{\hbar^{2}}}\nonumber\\
k_{2}&=&-\frac{m^{*}\alpha}{\hbar^{2}}+\sqrt{\left(\frac{m^{*}\alpha}{\hbar^{2}}\right)^{2}+\frac{2m^{*}E}{\hbar^{2}}}
\end{eqnarray}

The integrals in Eq.~(\ref{eq:Gr}) can be readily evaluated. The
term in $\widehat{G}_{\pm}(\vec{r}_{1},\vec{r}_{2},E)$ proportional
to the identity matrix is given by:\begin{eqnarray}
&&\lim_{\epsilon\rightarrow
0}\frac{-m^{*}}{(2\pi\hbar)^{2}}\int^{2\pi}_{0}\text{d}\phi\int^{\infty}_{0}\text{d}k
\left(\frac{k\exp(\text{i}kr\cos(\phi))}{(k-k_{1}\pm
\text{i}\epsilon)(k+k_{2})}+\frac{k\exp(\text{i}kr\cos(\phi))}{(k+k_{1})(k-k_{2}\pm\text{i}\epsilon)}\right)\widehat{1}\nonumber\\
&=&\lim_{\epsilon\rightarrow
0}\frac{-m^{*}}{2\pi\hbar^{2}}\int^{\infty}_{0}\text{d}k
\frac{J_{0}(kr)}{k_{1}+k_{2}}\left(\frac{k_{1}}{k+k_{1}}+\frac{k_{1}}{k-k_{1}\pm
\text{i}\epsilon}+\frac{k_{2}}{k+k_{2}}+\frac{k_{2}}{k-k_{2}\pm
\text{i}\epsilon}\right)\widehat{1}\nonumber\\
&=&-\frac{\text{i}m^{*}}{2\hbar^{2}}\left(\frac{k_{1}}{k_{1}+k_{2}}H^{\pm}_{0}(k_{1}r)+\frac{k_{2}}{k_{1}+k_{2}}H^{\pm}_{0}(k_{2}r)\right)\widehat{1}
\end{eqnarray}
which is just the weighted average of two free particle Green's
function with different wave vectors, $k_{1}$ and $k_{2}$.

Using the following two integrals:
\begin{eqnarray}
\int^{2\pi}_{0}\text{d}\phi\exp(\text{i}kr\cos(\phi))\sin(\phi)&=&0\nonumber\\
\int^{2\pi}_{0}\text{d}\phi\exp(\text{i}kr\cos(\phi))k\cos(\phi)&=&-\text{i}\frac{\partial}{\partial r}\int^{2\pi}_{0}\text{d}\phi\exp(\text{i}kr\cos(\phi))\nonumber\\
&=&-\text{i}\frac{\partial}{\partial r}2\pi J_{0}(kr)\nonumber\\
\end{eqnarray}

The terms in Eq.~(\ref{eq:Gr}) proportional to
$\widehat{\sigma}_{X}$ and $\widehat{\sigma}_{Y}$ are given
by\begin{eqnarray} &&\lim_{\epsilon\rightarrow
0}\frac{-m^{*}}{(2\pi\hbar)^{2}}\int^{2\pi}_{0}\text{d}\phi\int^{\infty}_{0}\text{d}k\frac{kZ^{\dagger}(\theta)(\widehat{\sigma}_{X}\cos(\phi)-\widehat{\sigma}_{Y}\sin(\phi))Z(\theta)}{(k-k_1)(k+k_2\pm\text{i}\epsilon)}\nonumber\\
&&-\frac{kZ^{\dagger}(\theta)(\widehat{\sigma}_{X}\cos(\phi)-\widehat{\sigma}_{Y}\sin(\phi))Z(\theta)}{(k+k_1\pm\text{i}\epsilon)(k-k_2)}\nonumber\\
&=&\lim_{\epsilon\rightarrow
0}\frac{m^{*}\text{i}}{2\pi\hbar^{2}}\widehat{Z}^{\dagger}(\theta)\widehat{\sigma}_{X}\widehat{Z}(\theta)\frac{\partial}{\partial
r}\int^{\infty}_{0}\text{d}k
\frac{J_{0}(kr)}{k_{1}+k_{2}}\left(\frac{1}{k+k_{1}}+\frac{1}{k-k_{1}\pm\text{i}\epsilon}-\frac{1}{k+k_{2}}-\frac{1}{k-k_{2}\pm\text{i}\epsilon}
\right)\nonumber\\
&=&\frac{-\widehat{Z}^{\dagger}(\theta)\widehat{\sigma}_{X}\widehat{Z}(\theta)m^{*}}{2(k_{1}+k_{2})\hbar^{2}}\frac{\partial}{\partial r}\left(H^{\pm}_{0}(k_{1}r)-H^{\pm}_{0}(k_{2}r)\right)\nonumber\\
&=&\frac{m^{*}}{2\hbar^{2}}\left(\widehat{\sigma}_{X}\cos(\theta)-\widehat{\sigma}_{Y}\sin(\theta)\right)\left(\frac{k_{1}}{k_{1}+k_{2}}H^{\pm}_{1}(k_{1}r)-\frac{k_{2}}{k_{1}+k_{2}}H^{\pm}_{1}(k_{2}r)\right)\end{eqnarray}

The Green's function,
$\widehat{G}_{\pm}(\vec{r}_{1},\vec{r}_{2},E)$, can finally be
written as
\begin{eqnarray}
\widehat{G}_{\pm}(\vec{r}_{1},\vec{r}_{2},E)&=&-\text{i}\frac{m^{*}}{2\hbar^{2}}\frac{k_{1}}{k_{1}+k_{2}}\left(\begin{array}{cc}H^{\pm}_{0}(k_{1}r)&\text{i}H^{\pm}_{1}(k_{1}r)\exp\left(\text{i}\theta\right)\\
\text{i}H_{1}^{\pm}(k_{1}r)\exp(-\text{i}\theta)&H_{0}^{\pm}(k_{1}r)\end{array}\right)\nonumber\\
&-&\text{i}\frac{m^{*}}{2\hbar^{2}}\frac{k_{2}}{k_{1}+k_{2}}\left(\begin{array}{cc}H^{\pm}_{0}(k_{2}r)&-\text{i}H^{\pm}_{1}(k_{2}r)\exp\left(\text{i}\theta\right)\\
-\text{i}H_{1}^{\pm}(k_{2}r)\exp(-\text{i}\theta)&H_{0}^{\pm}(k_{2}r)\end{array}\right)\nonumber\\
\label{eq:GF}
\end{eqnarray}
As noted earlier, $\widehat{G}_{\pm}(\vec{r}_{1},\vec{r}_{2},E)$ is
similar in form to $\widehat{G}^{k}_{0}(\vec{R})$ operator
[Eq.~(\ref{eq:GGG})] found in the partial wave scattering analysis
given in the text.  It is worth pointing out that, as was the case
for $\widehat{G}^{k}_{0}(\vec{R})$,
$\widehat{G}_{\pm}(\vec{r}_{1},\vec{r}_{2},E)\widehat{G}_{\pm}(\vec{r}_{2},\vec{r}_{1},E)\propto
\widehat{1}$, i.e., no net spin rotation is observed when the
particle traverses no net distance along a one-dimensional path.

Consider the experiment shown in Figure \ref{fig:figure2} for the
case when the detector and the emitter are one and the same.  With
the approximation to $\widehat{G}^{k}_{0}(\vec{R})$ made in
Eq.~(\ref{eq:GGGapprox}), it was concluded that Rashba spin-orbit
coupling does not give rise to modulations in the net current as a
function of $\vec{R}_{t}$ [Eq.~(\ref{eq:fluxsame})].  It is worth
pointing out that the conclusion reached by Eq.~(\ref{eq:fluxsame})
is valid, even if the approximation in Eq.~(\ref{eq:GGGapprox}) is
not made.  Treating the emitter as a point source, the scattered
wave function at the detector/emitter is given by
Eq.~(\ref{eq:swavey}):
\begin{eqnarray}
\Psi(\vec{r}_{d})&=&\sum_{k=1}^{N}\widehat{G}^{k}_{0}(\vec{r}_{d})\Psi(\vec{r}_{k})\nonumber\\
&=&\sum_{k=1}^{N}\widehat{G}^{k}_{0}(\vec{r}_{d})\widehat{G}_{+}(\vec{r}_{k},\vec{r}_{d},E)\widehat{\eta}
\end{eqnarray}
where $\widehat{\eta}$ represents the spin state of the injected
electron (can be taken to be either $\binom{1}{0}$ or $\binom{0}{1}$
or some linear combination of the two).  It is easy to show,
however, that
$\widehat{G}^{k}_{0}(\vec{r}_{d})\widehat{G}_{+}(\vec{r}_{k},\vec{r}_{d},E)$
is not proportional to the identity matrix.  For
$\overline{k}r_{k,d}\gg 1$,
$\widehat{G}^{k}_{0}(\vec{r}_{d})\widehat{G}_{+}(\vec{r}_{k},\vec{r}_{d},E)$
can be written as
\begin{eqnarray}
\widehat{G}^{k}_{0}(\vec{r}_{d})\widehat{G}_{+}(\vec{r}_{k},\vec{r}_{d},E)&\propto&\frac{\exp(\text{i}2\overline{k}r_{k,d})}{r_{k,d}}\left(\begin{array}{cc}a_{k}&b_{k}\exp(\text{i}\theta_{d}^{k})\\b_{k}\exp(-\text{i}\theta_{d}^{k})&a_{k}
\label{eq:noapprox}\end{array} \right)\end{eqnarray} where
\begin{eqnarray}
a_{k}&=&\sqrt{\frac{k_2}{k_1}}t^{1}_{k,0}+\sqrt{\frac{k_1}{k_2}}t^{2}_{k,0}\nonumber\\
b_{k}&=&\sqrt{\frac{k_1}{k_2}}t^{2}_{k,0}-\sqrt{\frac{k_2}{k_1}}t^{1}_{k,0}\nonumber\\
\end{eqnarray}
Eq.~(\ref{eq:noapprox}) indicates that the incident spin state,
$\widehat{\eta}$, is modified after following the effective 1D
trajectory (although Eq.~(\ref{eq:noapprox}) doesn't represent a
spin rotation since it is not unitary). However, the only distance
dependent factor in Eq.~(\ref{eq:noapprox}) contains
$\overline{k}r_{k,d}$ and does not depend upon $k_{\alpha}$.
Calculations of $\Delta\mu(\vec{R}_{t})$ therefore will not possess
any modulations which depend upon the Rashba interaction,
$k_{\alpha}$, supporting the original conclusions made in
Eq.~(\ref{eq:fluxsame}).


\begin{thebibliography}{10}

\bibitem{Wolf01}
S.A. Wolf, D.D. Awschalom, R.A. Buhrman, J.M. Daughton, S. von
Molnar, M.L.
  Roukes, A.Y. Chtchelkanova, and D.M. Treger, Science {\bf 294},  1488
  (2001).

\bibitem{Awschalombook}
{\em Semiconductor Spintronics and Quantum Computation}, edited by
D.D.
  Awschalom, D. Loss, and N. Samarth (Springer-Verlag, New York, 2002).

\bibitem{Zutic04}
I. Zutic, J. Fabian, and S.~Das Sarma, Rev. Mod. Phys. {\bf 76},
323  (2004).

\bibitem{Datta90}
S. Datta and B. Das, Appl. Phys. Lett. {\bf 56},  665  (1990).

\bibitem{Bychkov84}
Y.A. Bychkov and E.I. Rashba, J. Phys. C {\bf 17},  6039  (1984).

\bibitem{Dresselhaus55}
G. Dresselhaus, Phys. Rev. {\bf 100},  580  (1955).

\bibitem{Nitta97}
J. Nitta, T.A. Kazaki, H. Takayanagi, and T. Enoki, Phys. Rev. Lett.
{\bf 78},
  1335  (1997).

\bibitem{Mishchenko04}
E.G. Mishchenko, A.V. Shytov, and B.I. Halperin, Phys. Rev. Lett.
{\bf 93},
  226602  (2004).

\bibitem{Burkov04}
A.A. Burkov, A.S. Nunez, and A.H. MacDonald, Phys. Rev. B {\bf 70},
155308
  (2004).

\bibitem{Schliemann03}
J. Schliemann and D. Loss, Phys. Rev. B {\bf 68},  165311  (2003).

\bibitem{Hirsch99}
J.E. Hirsch, Phys. Rev. Lett. {\bf 83},  1834  (1999).

\bibitem{Zhang00}
S.F. Zhang, Phys. Rev. Lett. {\bf 85},  393  (2000).

\bibitem{Topinka01}
M.A. Topinka, B.J. LeRoy, R.M. Westervelt, S.E.J. Shaw, R.
Fleischmann, E.J.
  Heller, K.D. Maranowski, and A.C. Gossard, Nature {\bf 410},  183  (2001).

\bibitem{shawpap}
S.E.J. Shaw, R. Fleischmann, and E.J. Heller, Cond-mat/0105354
(2001).

\bibitem{Bulgakov01}
E.N. Bulgakov and A.F. Sadreev, JETP Lett. {\bf 73},  505  (2001).

\bibitem{Cserti04}
J. Cserti, A. Csordas, and U. Zulicke, Phys. Rev. B {\bf 70},
233307  (2004).

\bibitem{Herschthesis}
J.S. Hersch, Ph.D. thesis, Harvard University, 1999, chapter 6.

\bibitem{Inoue03}
J. Inoue, G.E.W. Bauer, and L.W. Molenkamp, Phys. Rev. B {\bf 67},
033104
  (2003).

\bibitem{Topinka00}
M.A. Topinka, B.J. LeRoy, S.E.J. Shaw, E.J. Heller, R.M. Westervelt,
K.D.
  Maranowski, and A.C. Gossard, Science {\bf 289},  2323  (2000).

\bibitem{Topinka03}
M.A. Topinka, R.M. Westervelt, and E.J. Heller, Phys. Today {\bf
56},  47
  (2003).

\bibitem{LeRoy05}
B.J. LeRoy, A.C. Bleszynski, K.E. Aidala, R.M. Westervelt, A.
Kalben, E.J.
  Heller, S.E.J. Shaw, K.D. Maranowski, and A.C. Gossard, Phys. Rev. Lett. {\bf
  94},  126801  (2005).

\bibitem{Miller03}
J.B. Miller, D.M. Zumbuhl, C.M. Marcus, Y.B. Lyanda-Geller, D.
  Goldhaber-Gordon, K. Campman, and A.C. Gossard, Phys. Rev. Lett. {\bf 90},
  076807  (2003).

\bibitem{Shawthesis}
S.E.J. Shaw, Ph.D. thesis, Harvard University, 2002.

\bibitem{Heller05}
E.J. Heller, K.E. Aidala, B.J. LeRoy, A.C. Bleszynski, A. Kalben,
R.M.
  Westervelt, K.D. Maranowski, and A.C. Gossard, Nano Lett. {\bf 5},  1285
  (2005).

\bibitem{Grundler00}
D. Grundler, Phys. Rev. Lett. {\bf 84},  6074  (2000).

\bibitem{Houten89}
H. van Houten, C.W.J. Beenakker, J.G. Williamson, M.E.I. Broekaart,
P.H.M. van
  Loosdrecht, B.J. van Wees, J.E. Mooij, C.T. Foxon, and J.J. Harris, Phys.
  Rev. B {\bf 39},  8556  (1989).

\bibitem{LandauSM1book}
L.D. Landau and E.M. Lifshitz, {\em Statistical Physics I}, 3rd  ed.
  (Butterworth-Heinemann, Oxford, 1999).

\bibitem{Elliott54}
R.J. Elliott, Phys. Rev. {\bf 96},  266  (1954).

\end{thebibliography}
\end{document}